\documentclass[preprint,amsmath,showpacs,nofootinbib,showkeys]{revtex4-1}
\usepackage[T1]{fontenc} 
\usepackage{multirow}
\usepackage{makecell}
\usepackage{longtable}
\usepackage{appendix}
\usepackage{ulem}

\newcommand{\kev}{\ensuremath{\,\mathrm{keV}}}
\newcommand{\mev}{\ensuremath{\,\mathrm{MeV}}}
\newcommand{\gev}{\ensuremath{\,\mathrm{GeV}}}
\newcommand{\tev}{\ensuremath{\,\mathrm{TeV}}}

\usepackage{amsmath}
\usepackage{graphicx,bm}
\usepackage[colorlinks,citecolor=blue]{hyperref}
\usepackage[caption=false]{subfig}
\usepackage{slashed}
\usepackage{ulem}

\begin{document}
\title{Inelastic Scattering of Dark Matter with Heavy Cosmic Rays}

\author{Keyu Lu$^{a}$}
\author{Yue-Lin Sming Tsai$^{b,c}$}\thanks{smingtsai@pmo.ac.cn}
\author{Qiang Yuan$^{b,c}$}\thanks{yuanq@pmo.ac.cn}
\author{Le Zhang$^{a,d,e}$}\thanks{zhangle7@mail.sysu.edu.cn}

\affiliation{$^{a}$School of Physics and Astronomy, Sun Yat-sen University, Zhuhai, 519082, P. R. China}
\affiliation{$^{b}$Key Laboratory of Dark Matter and Space Astronomy, 
Purple Mountain Observatory, Chinese Academy of Sciences, Nanjing 210023, P. R. China}
\affiliation{$^{c}$School of Astronomy and Space Science, University of Science and Technology of China, Hefei, Anhui 230026, P. R. China}
\affiliation{$^{d}$CSST Science Center for the Guangdong-Hong Kong-Macau Greater Bay Area, SYSU, Zhuhai 519082, P. R. China}
\affiliation{$^{e}$Peng Cheng Laboratory, No. 2, Xingke 1st Street, Shenzhen 518000, P. R. China}

\begin{abstract}
We investigate the impact of inelastic collisions between dark matter (DM) and heavy cosmic ray (CR) nuclei on CR propagation. We approximate the fragmentation cross-sections for DM-CR collisions using collider-measured proton-nuclei scattering cross-sections, allowing us to assess how these collisions affect the spectra of CR Boron and Carbon. We derive new CR spectra from DM-CR collisions by incorporating their cross-sections into the source terms and solving the diffusion equation for the complete network of reactions involved in generating secondary species. 
In a specific example with a coupling strength of $b_{\chi}=0.1$ and a DM mass of $m_{\chi}=0.1~\gev$, considering a simplified scenario where DM interacts exclusively with Oxygen, a notable modification in the Boron-to-Carbon spectrum due to the DM-CR interaction is observed. Particularly, the peak within the spectrum, spanning from $0.1~\gev$ to $10~\gev$,  experiences an enhancement of approximately 1.5 times. However, in a more realistic scenario where DM particles interact with all CRs, this peak can be amplified to twice its original value.
Utilizing the latest data from AMS-02 and DAMPE on the Boron-to-Carbon ratio, we estimate a 95\% upper limit for the effective inelastic cross-section of DM-proton as a function of DM mass. Our findings reveal that at $m_\chi\simeq 2\mev$, the effective inelastic cross-section between DM and protons must be less than $\mathcal{O}(10^{-32})~{\rm cm}^2$.
\end{abstract}

\date{\today}

\maketitle

\tableofcontents

\newpage
\section{Introduction} 

Dark matter (DM) constitutes the majority of the matter in the universe and 
is thought to form a halo around our Milky Way comprised of non-relativistic particles. 
However, DM has only been observed through its gravitational interactions with standard model matter, such as gravitational lensing~\cite{astro-ph/0608407}. 
To detect non-gravitational interactions, several methods have been developed, including 
DM direct detection~\cite{1811.11241,2007.08796,2007.15469,2107.13438,2112.03920,2207.03764,2207.11330}, 
DM indirect detection~\cite{astro-ph/0310473,astro-ph/0603425,0704.2444,0807.3429,0905.4952,0902.1089,0906.1197,0912.4504, 1503.04858, 1503.02641, 1705.02358, 1803.09734, 1809.08610, 1908.02668,2006.00513,2007.15006,2008.11561,2205.12950,2307.04130}, and accelerator detection~\cite{2107.13021,2108.13391}. 
Those instruments focus primarily on detecting DM with a mass heavier than the GeV scale. 
If DM particles are lighter than GeV, they can evade the direct detection detector energy threshold 
due to their non-relativistic velocity.

Cosmic Rays (CRs) are high-energy particles originating from various astrophysical sources. 
In the event of a collision between high-energy CR particles and DM particles, 
the energy transfer between the two can potentially accelerate DM particles to 
a higher energy range~\cite{1810.10543, 1811.00520, 1912.09904, 1906.11283, 2004.03161, 2005.09480, 2009.00353, 2110.08863,2111.05559, 2210.01815, 2211.13399, 2212.02286, 2112.08957, 2305.12668, 2307.09460, 2308.02204, 2309.11003}. 
Alternatively, such collisions may also produce an indirect signal 
through smashed CRs~\cite{astro-ph/0203240,1802.03025,1810.00372,2002.11732,2008.12137}. 
Therefore, incorporating CR-DM collisions in our analysis can enable us to explore sub-GeV DM parameter space 
within the current sensitivity of current detectors.

Although the fragmentation of heavy nuclear is well-studied in CR physics, see e.g., Ref.~\cite{1803.04686}, 
the mechanism of the inelastic scattering between DM and CR heavy nuclei is still unclear~\cite{2108.00583,2209.03360,2210.05685}. 
Therefore, we make two hypotheses to derive the DM-nucleus inelastic interaction cross-section:
\begin{itemize}
    \item When considering two collisions DM-CR and proton-CR, 
    if the same kinetic energy of DM and proton are observed in the CR rest frame, 
    the final kinetic energy distributions of particles in DM-CR and proton-CR are identical. 
    
    \item The DM-CR cross-section $\sigma_{\chi-{\rm CR}}$ and the proton-CR cross-section $\sigma_{p-{\rm CR}}$ 
    are related by a phenomenological constant factor $b_\chi$, namely $b_\chi\equiv\sigma_{\chi-{\rm CR}}/\sigma_{p-{\rm CR}}$.
\end{itemize} 

Thanks to the recent development of satellite telescopes such as AMS-02~\cite{AMS:2016brs} and DAMPE~\cite{1706.08453}, 
the statistical uncertainties of CR fluxes, especially for those secondary particles, have significantly improved. 
In this context, CR secondary particles may provide a useful probe for detecting inelastic scattering between DM and CRs. 
In this work, we propose to identify the DM-inelastic scattering between DM and CR heavy nuclei from 
the Carbon and Boron measured by AMS02~\cite{AMS:2016brs} and DAMPE~\cite{2210.08833}.

The systematic uncertainties, including the diffusion coefficients $D_0$ and heights of the diffusion zone $Z_h$ may dilute the impact of the inelastic collisions between DM and heavy CR nuclei on the CR propagation.  
To figure out their impacts on the $95\%$ upper limits of $b_\chi$, 
we simulate the DM-induced CR spectra with 
the $1\sigma$ favored region of $D_0$ and $Z_h$ given by Ref.~\cite{1810.03141}, 
despite of uncertainty degeneracy. 
Our work reveals that when considering uncertainties, the upper limits of $b_\chi$ for dark matter masses greater than $10\mev$ can be weakened by more than one order of magnitude.
However, for DM masses lighter than $1\mev$, DM signal can only contribute to the high-energy spectrum where the experimental error bars are significantly larger than those in the low-energy spectrum. 
The upper limits of $b_\chi$ with $m_\chi\approx 0.1\mev$ can only be slightly altered if including $D_0$ and $Z_h$ systematic uncertainties.

The remainder of this paper is organized as follows. 
At the beginning of Sec.~\ref{sec:frame}, we introduce the standard framework for CR propagation and 
discuss the propagation equation incorporating the inelastic scattering between DM and CRs. 
Then, in Sec.~\ref{sec:Xsec}, we calculate the $\chi-$CRs inelastic scattering cross-section 
by utilizing currently available collider data. 
Subsequently, in Sec.~\ref{sec:Numerical}, we simulate the energy spectra of Carbon C, Oxygen O, Boron B, 
and Boron-to-Carbon ratio (B/C) for cosmic rays with different DM particle masses.
We also evaluate the characteristic signals of $\chi-$CRs interaction. 
Finally, in Sec.~\ref{sec:upperlimits}, we show the exclusion region of $\chi-$CRs interaction 
with the DM mass range between $10^{-4}\gev$ and $10^2\gev$ by fitting the Boron-to-Carbon ratio.

\section{Cosmic rays propagation in the presence of DM}
\label{sec:frame}

It is known that charged CRs diffuse in a random magnetic field of the Milky Way
and collide with the interstellar medium (ISM) gas.
During the propagation of cosmic nuclei, the spallation of cosmic nuclei can take place due to 
the collision. CR particles gain or lose energy and fragment into secondary particles. 
Naively, DM can also collide with high-energy cosmic nuclei to 
smash nuclei. 
Hence, we can modify the diffusion equation of CRs in the Milky Way~\cite{astro-ph/0701517,0807.4730} by including CRs-DM collisions as  
\begin{equation}
\begin{aligned}
\frac{\partial N_i(p,r,z)}{\partial t} &-\boldsymbol{\nabla} \cdot\left(D_{xx} \boldsymbol{\nabla}-\boldsymbol{v}_c\right) N_i+\frac{\partial}{\partial p}\left(\dot{p}-\frac{p}{3} \boldsymbol{\nabla} \cdot \boldsymbol{v}_c\right) N_i-\frac{\partial}{\partial p} p^2 D_{p p} \frac{\partial}{\partial p} \frac{N_i}{p^2}\\
&=Q_i(p, r,z )
+\sum_{k=\rm{DM,~gas}}\left[
\sum_{\mathcal{A}_j>\mathcal{A}_i} \Gamma_{k}^{s}(j \to i; T) 
-\frac{N_i}{\tau_{k}^{f,i}(T)}
\right]
-\frac{N_i}{\tau_i^{r}}\,.
\end{aligned}
\label{eq:cr-prop}
\end{equation}
Here, $N_i(p,r,z)$ refers to the differential number density of the $i$-th atomic species of CRs per unit momentum interval, and 
it is a function of the particle momentum $p$ and the position in cylindrical coordinates $(r,z)$. 
The convection velocity and the momentum loss rate are represented by $\boldsymbol{v}_c$ and $\dot{p}\equiv{\rm d}p/{\rm d}t$, respectively. 
For the purposes of our analysis, we assume a spatially homogeneous diffusion coefficient $D_{xx}$, as described in Ref.~\cite{1001.0553},
\begin{equation}
D_{xx} = D_0\beta^\eta \left( R/R_0 \right)^{\delta}, 
\end{equation} 
where a particle with charge $Ze$ has the rigidity $R\equiv pc/Ze$ 
and $D_0$ is a normalization parameter. The velocity of the particle, represented by $\beta$, is measured in units of the speed of light $c$. 
The power index $\delta$ reflects the property of the interstellar medium (ISM) turbulence, and a value of $\delta= 1/3$ for a Kolmogorov spectrum of turbulence is taken. 
$\mathcal{R}_0\equiv 4~\mathrm{GV}$ is a reference rigidity
To improve the fit of the data, a phenomenological modification of the diffusion coefficient at low energies is introduced through the parameter $\eta$, as discussed in~\cite{0909.4548}. 

The process of reaccelerating CRs during their propagation in the turbulent galactic magnetic field is described by the momentum diffusion term $D_{pp}$. As shown in~\cite{1994ApJ...431..705S}, the diffusion coefficient $D_{pp}$ can be related to $D_{xx}$ via 
\begin{equation}
D_{pp}D_{xx}=\frac{4p^2v_A^2}{3\delta(4-\delta^2)(4-\delta)},
\end{equation}
where $v_A$ is the Alfven speed.

On the right-hand side of Eq.~\eqref{eq:cr-prop}, we incorporate all the interaction terms 
between CRs and ISM gas, as well as between CRs and DM. Note that the CR source term, except for spallation, is contained within $Q_i(p,r,z)$.  
We parameterize the time scales for fragmentation and radioactive decay as $\tau^f$ and $\tau^r$, respectively. 
The kinetic energy per nucleon is $T \equiv(E-m_A)/\mathcal{A}$ with the total energy of a nucleus $E$, mass number $\mathcal{A}$, 
and mass $m_A \simeq \mathcal{A} m_{\rm nucleon}$. 
In the most general form, the CR fragmentation due to gas or DM ($k=$ DM, gas) can be written as 
\begin{equation}\label{eq:frag}
\frac{1}{\tau_k^{f,i}(T)}\equiv 
n_{k}\sigma_{k, i}(T) \beta^i(T) c \,,
\end{equation}
where $k$ runs over DM, Hydrogen (H), and Helium (He), 
while $i$ indicates the $i$-th CR atomic species. 
The CR-DM, CR-Hydrogen, and CR-Helium cross sections are counted as $\sigma_{k,i}$.  
The total interstellar hydrogen density is $n_{\rm H}=n_{\rm HI}+2 n_{{\rm H}_2}+n_{\rm HII}$ and $n_{\rm He}\simeq 0.11 n_{\rm H}$. 
In this work, we safely ignore the contribution of heavier elements of the gas as CR targets.
This is a reduction term, thus a negative sign in front of the CR fragmentation term.  
On the other hand, the production of secondary CR $i$ generated from heavier element $j$ spallation can be described as
\begin{equation}
 {\rm CR}~j + k \rightarrow {\rm CR}~i + k + {\rm etc.}\,,
\end{equation}
and the total inelastic scattering rate summing over the atom number condition $\mathcal{A}_j>\mathcal{A}_i$ cases is 
\begin{equation}\label{eq:sec}
\sum_{\mathcal{A}_j>\mathcal{A}_i} \Gamma_{k}^{s}(j \to i; T)=cn_{k} \sum_{\mathcal{A}_j>\mathcal{A}_i}   
\int d T^{\prime} \beta\left(T^{\prime}\right) N_j\left(T^{\prime}\right)
\left[\frac{d \sigma_{k}}{d T}\left(j \rightarrow i; T, T^{\prime}\right)
\right]\,,
\end{equation}
where $T^\prime$ is the kinetic energy of the heavier CR particle $j$. 
The differential cross section of ${\rm CR}~j-k$ inelastic scattering is $d\sigma_{k}/dT$. 
The calculations for $d\sigma_{k={\rm H, He}}/dT$ has been included in the code~\texttt{GALPROP}~\cite{astro-ph/9807150}.  
In this study, we use \texttt{GALPROP} to perform numerical calculations for the propagation of CRs.

\subsection{Collision cross section between CRs and DM}
\label{sec:Xsec}

Owing to our poor understanding of DM-nucleon interactions, the most straightforward method to simulate the fragmentation cross sections for all $\chi-$CRs inelastic collisions is to replicate the spectra of CRs and proton ($p$) inelastic collisions. This study is based on two assumptions:
\begin{enumerate}
    \item[(1)] Once the equivalent incoming kinetic energy of $\chi$ and $p$ is observed in the CR rest frame, the final particle kinetic energy distributions in $\chi-$CRs and $p-$CR are identical.

    \item[(2)] The cross section for $\chi-$CRs can be obtained simply by scaling the cross section for $p$-CR through  
    $\sigma_{\chi-{\rm CR}}=b_\chi \cdot \sigma_{p-{\rm CR}}$. Here $b_\chi$ is a phenomenological constant factor that accounts for the strength of interactions between DM particles and cosmic rays.
\end{enumerate}

When the mass of the DM particle is lighter than that of the proton, the shifted spectrum lies below the binding energy. Therefore, we can set the binding energy ($B$) as the energy threshold by  utilizing the Bethe-Weizsäcker semi-empirical mass formula~\cite{1935ZPhy...96..431W},  
\begin{equation}
    \begin{aligned}
{ }_Z^\mathcal{A} B= & a_v \mathcal{A}-a_s \mathcal{A}^{2 / 3}-
a_c \frac{Z^2}{\mathcal{A}^{1 / 3}}-a_a \frac{(\mathcal{A}-2 Z)^2}{\mathcal{A}} 
 +{\rm sign}\times a_p \frac{1}{\mathcal{A}^{1 / 2}}.
\end{aligned}
\label{eq:binding}
\end{equation}

The variables $\mathcal{A}$ and Z represent, once again, the nucleon number and proton number, respectively. The coefficients $a_v$, $a_s$, $a_c$, $a_a$, and $a_p$ correspond to the volume, surface, Coulomb, asymmetry, and pair terms, respectively, as described in~\cite{KIRSON200829}. 
The values used in this study are as follows from~\cite{Benzaid:2020prt}: 
$a_v=14.9297\mev$, 
$a_s=15.0580\mev$, 
$a_c=0.6615\mev$, 
$a_a=21.6091\mev$, and 
$a_p=10.1744\mev$. 
The sign in front of $a_p$ is positive for an odd number of $\mathcal{A}$ and negative for an even number.

Next, we determine the mapping of kinetic energy in the case where the mass of DM particles differs from that of protons, i.e., $m_\chi\ne m_p$. Specifically, we consider the kinetic energies of DM and protons, denoted as $T^{[0]}_{\chi}$ and $T^{[0]}_{p}$, respectively, in the rest frame of the CR nucleus. Furthermore, we examine the kinetic energies of the CR nucleus, denoted as $T^{[\chi-{\rm Lab}]}_{\rm CR}$ and $T_{\rm CR}^{[p-{\rm Lab}]}$, respectively in the lab frame of $\chi-$CRs and $p-$CR. 
By using the four-momentum conservation and assuming the relationship $T^{[0]}_{\chi}=T^{[0]}_p$, 
we can derive the corresponding replacement, 
\begin{equation}\label{eq:Tlab}
T^{[p{\rm -Lab}]}_{\rm CR}\to \frac{m_\chi}{m_p} T^{[\chi{\rm -Lab}]}_{\rm CR}\,.
\end{equation}
Hence, the inelastic scattering cross section of $\chi-$CRs in Eqs.~\eqref{eq:frag} and \eqref{eq:sec} can be expressed as follows:
\begin{equation}\label{eq:dsigma/dT}
\frac{d\sigma_{\chi-{\rm CR}}}{dT}
\left(T=T^{[\chi-{\rm Lab}]}_{\rm CR}\right)=
b_{\chi} \frac{d\sigma_{p-{\rm CR}}}{dT}
\left(T=\frac{m_\chi}{m_p}  T^{[\chi-{\rm Lab}]}_{\rm CR}\right)\,.
\end{equation} 
We can obtain the $\chi-$CRs differential cross section $d\sigma_{\chi-{\rm CR}}/dT$ straightforwardly, 
by shifting the DM-proton differential cross section $d\sigma_{p-{\rm CR}}/dT$ 
towards lower (in the case of $m_\chi> m_p$) or higher (in the case of $m_\chi< m_p$) energies.

\subsection{Configuration of source and propagation parameters}
In the stationary limit, $\frac{\partial N_i}{\partial t} = 0$, we solve the propagation equation using the publicly available code \texttt{GALPROP}~\cite{astro-ph/9807150}. The boundary condition, $N_i(r,z = |L|) = 0$, is applied to simulate free particle escape at the Galactic boundaries. 
For illustrative purposes, we engage two benchmark propagation models, detailed in Table~\ref{tab:diff-para}. 
Model A is characterized by parameters that accurately predict the nuclear spectra, for Oxygen, Carbon, and Boron, as observed by AMS-02 and Voyager-1. Additionally, to accommodate both AMS-02 and DAMPE observations, we refine the propagation model to include an extra break in the diffusion coefficient~\cite{2210.09205}, suggesting a transition in the rigidity-dependence slope to $\delta_h$ for $\mathcal{R} > \mathcal{R}_h$ (Model B).


\begin{table}
\begin{center}
\begin{tabular}{>{\centering\arraybackslash}p{0.7\textwidth}|>{\centering\arraybackslash}p{0.1\textwidth}|>{\centering\arraybackslash}p{0.1\textwidth}}
\hline
\textbf{Model} & \textbf{A} & \textbf{B} \\
\hline\hline
Diffusion coefficient $D_0~\left(10^{28}~{\rm cm^2/s}\right)$  &  4.10 &  3.32\\
Diffusion coefficient rigidity index $\delta$& 0.477  & 0.60\\
Diffusion coefficient velocity index $\eta$ & $-1.51$ &  $-0.61$\\
Gradient of convection velocity $dV_c/dz~\left(\mathrm{km}\,\mathrm{s}^{-1}\,\mathrm{kpc}^{-1}\right)$& 0.0 & 0.0\\
Alfv\'en speed $v_A~\left(\mathrm{km}~\mathrm{s}^{-1}\right)$&  --- & 22.4\\
Height of diffusion zone $Z_h~\left(\mathrm{kpc}\right)$&  4.93   & 3.61\\
Extra reference rigidity~$\mathcal{R}_h~\left(\mathrm{GV}\right)$& --- &212.5\\
Extra diffusion coefficient rigidity index~$\delta_h$& --- &0.25\\
Injection spectral index $\nu_0$&0.70&0.41\\
Injection spectral index $\nu_1$&2.40&2.35\\
Injection spectral index $\nu_2$& --- &2.42\\
Reference injection rigidity1 $\mathcal{R}_{br,1}~\left(\mathrm{GV}\right)$&1.31&1.02\\
Reference injection rigidity2 $\mathcal{R}_{br,2}~\left(\mathrm{GV}\right)$& --- &142.3\\
\hline
Solar modulation $\Phi~\left(\mathrm{GV}\right)$ & 0.742 & 0.690\\
\hline\hline
\end{tabular}
\end{center}
\caption{Fiducial propagation model utilized in this study. Detailed parameter estimation for 
Model A is available in Ref.~\cite{1810.03141,1701.06149}, from fitting to the AMS-02 data, 
and for Model B in Ref.~\cite{2210.09205} after adding the DAMPE B/C data.  
}
\label{tab:diff-para}
\end{table}

The primary CR source was described as $Q_i(p,r,z)$ in Eq.~\eqref{eq:cr-prop}, which can be rewritten as 
\begin{equation}
Q_i(p,r,z)=f(r,z) q_i(p)\,,
\end{equation}
where $f(r,z)$ and $q_i$ represent the spatial distribution and the injection spectrum of the CR nuclei source, respectively.

The nuclei injection spectrum is assumed to be a broken power-law function of rigidity
\begin{equation}
    q(\mathcal{R})=q_0 \mathcal{R}^{-\nu_0} \prod_{i=1}^n\left[1+\left(\frac{\mathcal{R}}{\mathcal{R}_{\mathrm{br}, i}}\right)^s\right]^{\left(\nu_{i-1}-\nu_i\right) / s}\,,
\end{equation}
where $\nu_0$ is the spectral index at the lowest energies, $\nu_{i-1}$ and $\nu_i$ are spectral indices below and above break rigidity $\mathcal{R}_{br,i}$, 
and $s$ describes the smoothness of the break which was fixed to be $s = 2$ throughout this work. 
We take $n = 2$ for Model B, and $n = 1$ for Model A.
The spatial distribution of the primary CR particles $f(r,z)$ 
is similar to that of supernova remnants (SNRs) as
\begin{equation}
f(r, z) =\left(\frac{r}{R_{\odot}}\right)^a \exp \left[-\frac{b\left(r-R_{\odot}\right)}{R_{\odot}}\right] \exp \left(-\frac{|z|}{z_s}\right)
\end{equation}
where $R_{\odot}=8.5$ kpc is the distance between the solar system and the Galactic center, 
and $z_s \approx 0.2$ kpc is the characteristic height of the Galactic disk. 
We choose the shape parameters $a=1.25$ and $b=3.56$ measured from~\cite{1011.0037} to match the Galactic diffuse $\gamma$-ray emission and the ratio of H$_2$ to CO.

In the case of $\chi-$CRs interactions, the density profile of the Milky Way halo is adopted to be the Navarro-Frenk-White (NFW) distribution~\cite{astro-ph/9508025}, expressed as:
\begin{equation}
\rho_\chi(r)=\frac{\rho_s}{\left(r / r_s\right)\left(1+r / r_s\right)^2},
\end{equation}
with $r_s=20$ kpc and $\rho_s=0.26$ GeV/${\rm cm}^3$, thus in agreement with the local density of $\rho=0.3$ GeV/${\rm cm^3}$ at $r=8.5$ kpc.

\section{Numerical results and discussion} 
\label{sec:Numerical}

Given the impact of interactions between DM and CRs on the spallation process of heavy elements, 
it is challenging to comprehensively analyze their collective collisions if considering all DM-CR interactions simultaneously. 
To address this issue, we introduce a simplified toy model in Sec.~\ref{sec:toy}, 
where DM particles exclusively interact with Oxygen. 
In this work, we fix the solar modulation to $\Phi=0.742$ GV for Model A and 0.69 GV for Model B~\footnote{
Although the solar modulation effect could potentially distort certain spectral characteristics,
particularly those in the lower energy range, we follow fix $\Phi_A=0.742$ GV for Model A
\cite{1810.03141}, and $\Phi_B=0.69$ GV for Model B \cite{2210.09205}}.
Subsequently, in Sec.\ref{sec:gen}, we present the complete scenario, 
followed by an examination of the systematic uncertainties from $D_0$ and $Z_h$ on CR spectra in Sec.\ref{sec:astro_unc}. 
Finally, we constrain the DM model parameters 
by employing the measured B/C ratio spectra from the AMS-02~\cite{AMS:2017seo,AMS:2018tbl} and DAMPE~\cite{1706.08453} in Sec.~\ref{sec:upperlimits}.

We find that the collision terms between CRs and DM particles in Eq.~\eqref{eq:cr-prop} 
are directly proportional to the DM parameters $b_\chi \rho_\chi/m_\chi$. 
While $b_\chi$ and $m_\chi$ are subject to a parameter-degeneracy, 
the peak of the DM-CR cross-section is also determined by $m_\chi$. 
To disentangle the scaling factor and kinematics in the collision terms, 
we introduce a phenomenological variable, $b_\chi/m_\chi$. 
This allows us to set a clear separation between the role played by each of these parameters.

In this study, we employ the chi-squared $\chi^2$ minimization to evaluate the statistical strength and   
constrain the DM collisions. 
We include the B/C ratio data from AMS-02 and DAMPE, and the total $\chi^2$ can be written as  
\begin{equation}
\chi^2\left(D_0, Z_h, m_\chi, b_\chi\right)=\sum_{\rm exper.} 
\sum_{i} \frac{1}{\sigma_i^2}\left( f^{\rm obs}_i- f^{\rm pred}_i\right)^2\,.
\label{eq:chisq}
\end{equation}
For each energy bin $i$, the experimental error is given by $\sigma_i$, while 
the predictions and observations of B/C ratio are denoted as $f_i^{\rm pred}$ and $f_i^{\rm obs}$. 
Summing two experiments, the total 80 data points of all measured B/C ratios 
are within the range $0.6\gev<E_k/n<4\tev$.

\subsection{A toy model: only DM-Oxygen collisions}
\label{sec:toy}

\begin{figure}
\centering
\includegraphics[width=0.32\linewidth]{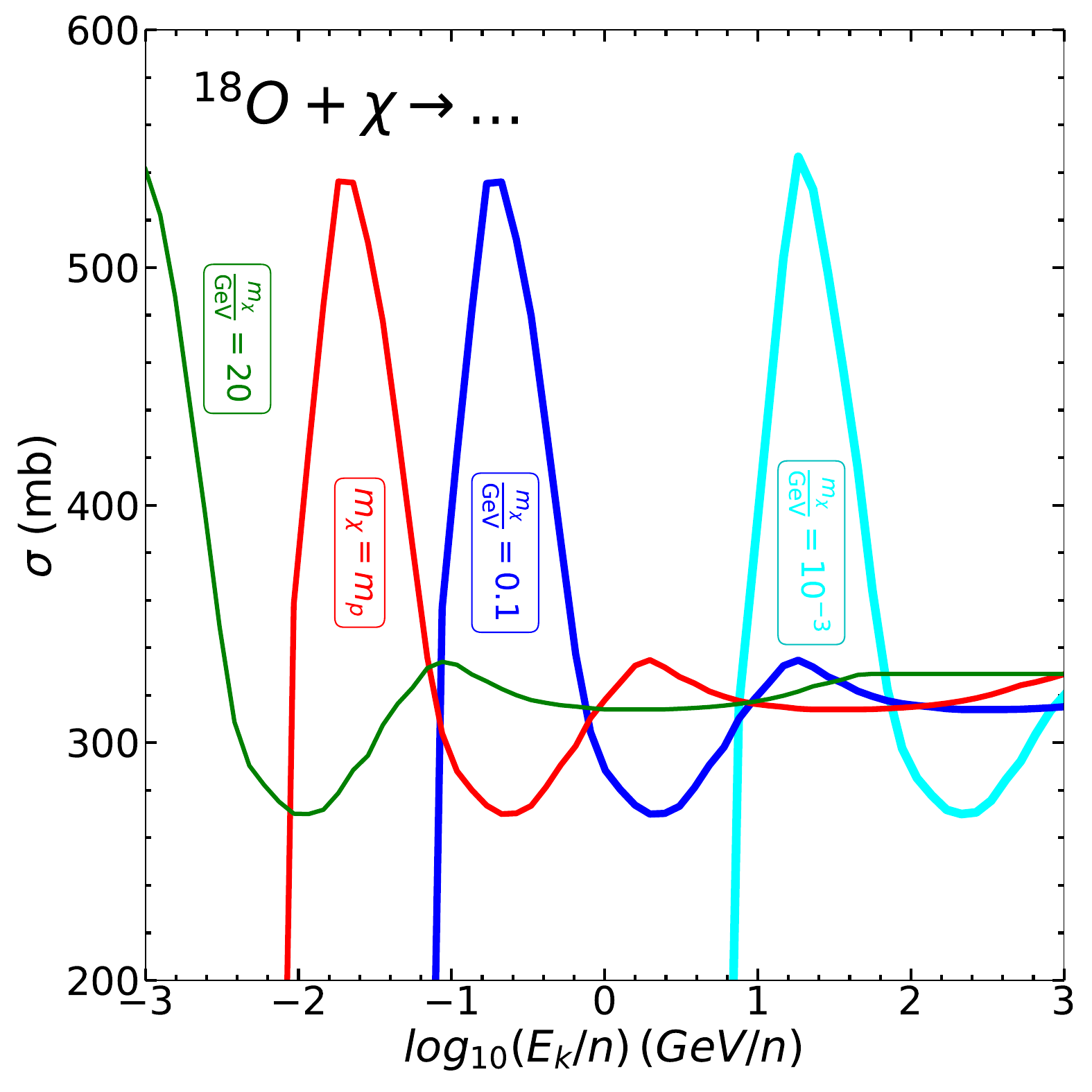}
\includegraphics[width=0.32\linewidth]{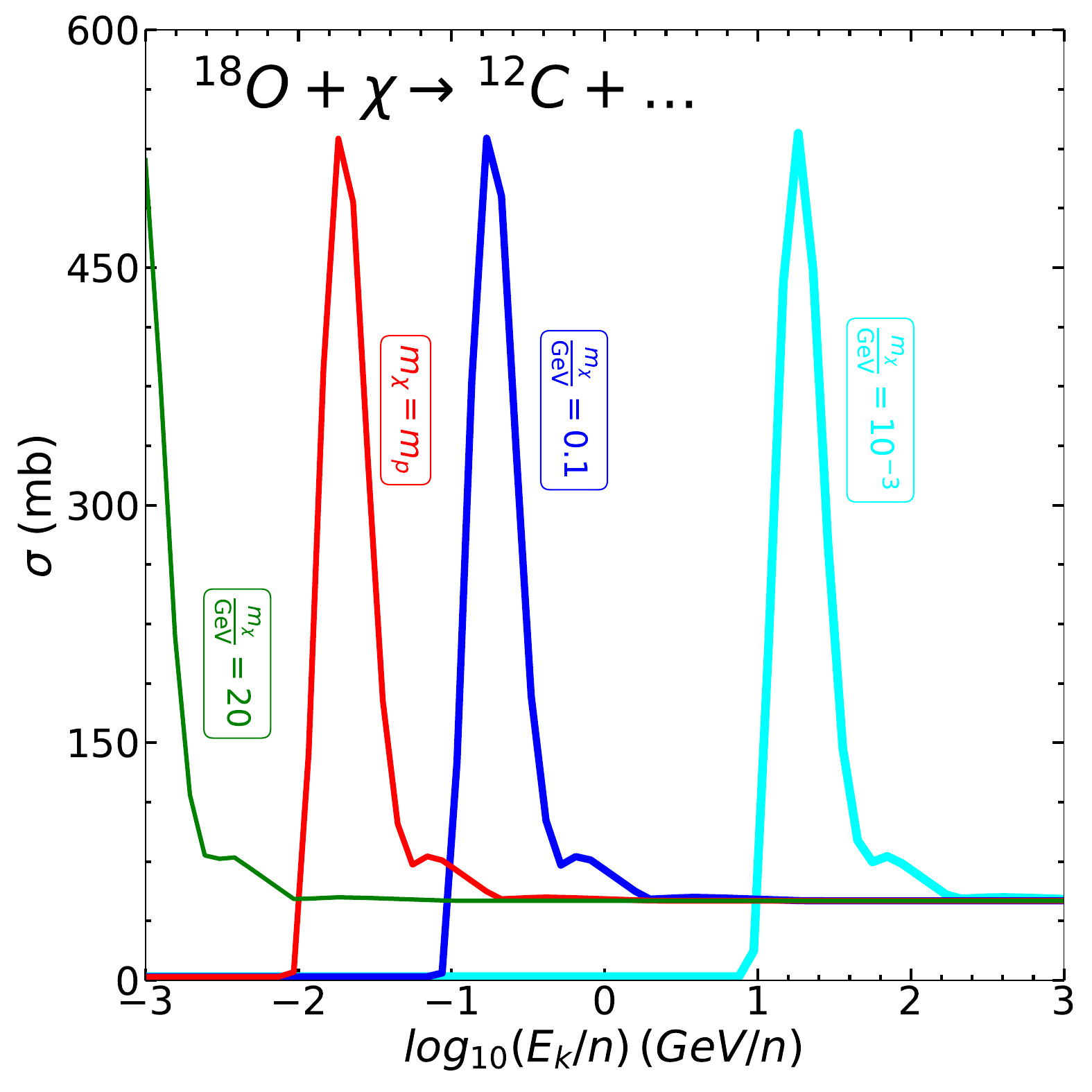}
\includegraphics[width=0.32\linewidth]{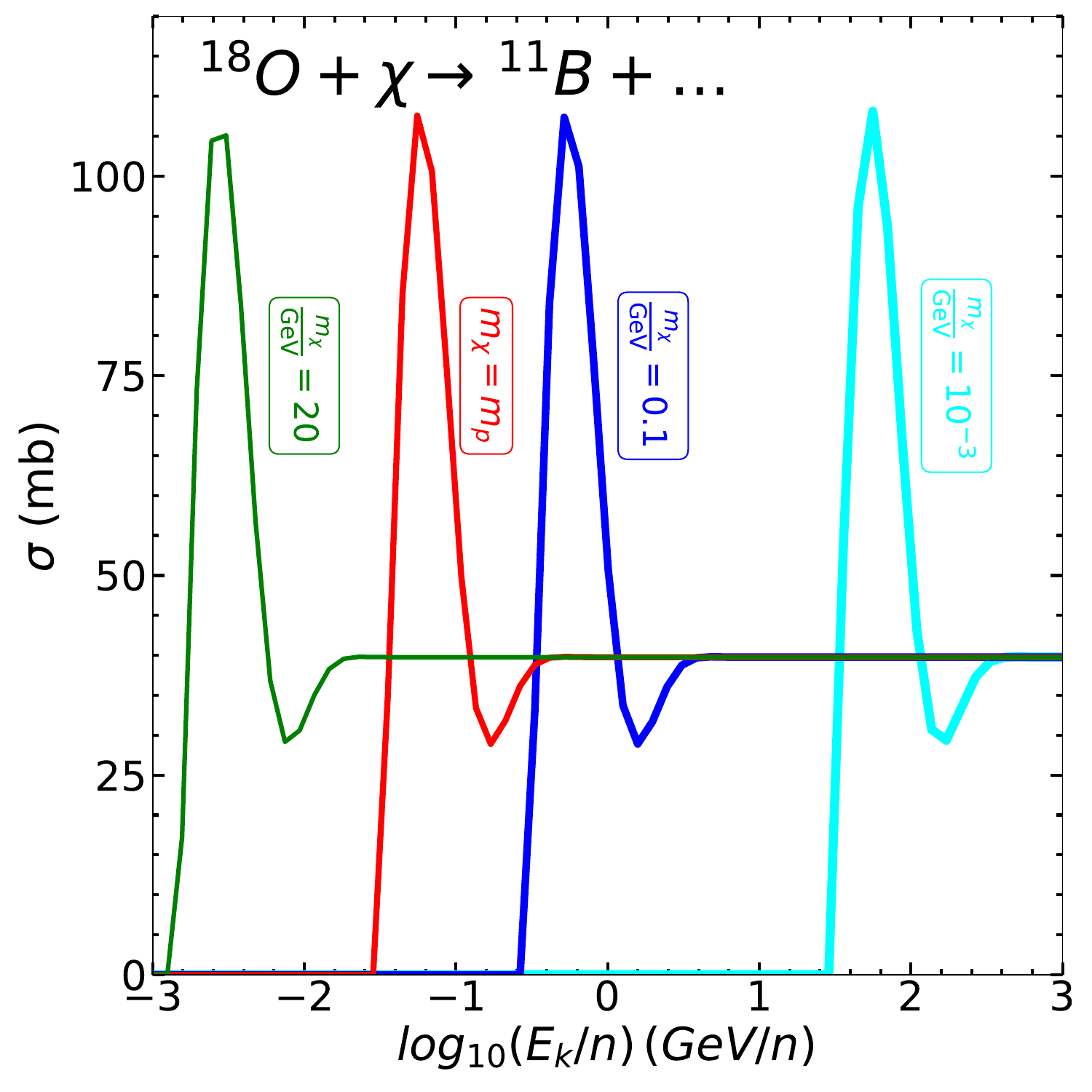}
    \caption{Comparison of the total cross-section for the fragmentation of Oxygen into all feasible lighter nuclei (left panel) 
    with the production cross-sections of Carbon (middle panel) and Boron (right panel), as a function of the kinetic energy per nucleon of $^{16}$O. 
    Four different DM benchmark masses $m_\chi=10^{-3}~\gev,\,0.1~\gev,\, m_p,\, 20\gev$ are presented by 
    cyan, blue, red, and green lines, respectively. 
    The strength of DM-CRs interactions is fixed to be $b_\chi=1$.
    }
    \label{fig:xsec}
\end{figure}

In this subsection, we consider a toy model that \textit{DM particles only interact with Oxygen}. 
This toy model is useful in tracking the cascades of elements with $Z\leq 8$ produced by DM-Oxygen collisions. 
The 
CR Oxygen abundance can be depleted not only by collisions with protons but also by collisions with DM particles. 
The left panel of Fig.~\ref{fig:xsec} displays the dependence of the total cross-section 
for the fragmentation of Oxygen into all possible lighter nuclei with respect to its kinetic energy per nucleon $E_k/n$.  
The production cross sections of Carbon (via $^{16}$O+ $\chi \to ^{12}$C+$\cdots$) and 
Boron (via $^{16}$O+ $\chi \to$$^{11}$B+$\cdots$) are illustrated in the middle and right panels, respectively. 
To demonstrate the kinematics of the collisions between DM and CRs, 
we take four representative DM masses: 
two lines (cyan and blue) have DM mass lighter than gas particles 
($10^{-3}\gev$ and $0.1\gev$), the red lines show the case $m_\chi=m_p$, 
and the green lines are with DM mass heavier than gas particles ($m_\chi=20\gev$). 
By taking $b_\chi/m_\chi=1\gev^{-1}$, the cross-section for $m_\chi=m_p$ case is exactly the same as the one for $^{16}$O-proton collisions.

Based on Fig.~\ref{fig:xsec}, our findings are summarized as follows.
\begin{itemize}
    \item The total fragmentation cross-section or the production cross-section of $^{10}$B or $^{12}$C exhibits a sharp peak. 
    \item  If increasing $m_\chi$, the peaks of the fragmentation and production cross sections shift to lower kinetic energy. 
    This is because of the factor $m_\chi/m_p$ in Eq.~\ref{eq:Tlab}. 
    \item  The $^{10}$B production cross section is one order of magnitude lower than $^{12}$C, 
    thus the primary product of $^{16}$O fragmentation is from Carbon.
\end{itemize}

\begin{figure}[ht!]
    \centering
    \includegraphics[width=0.496\linewidth]{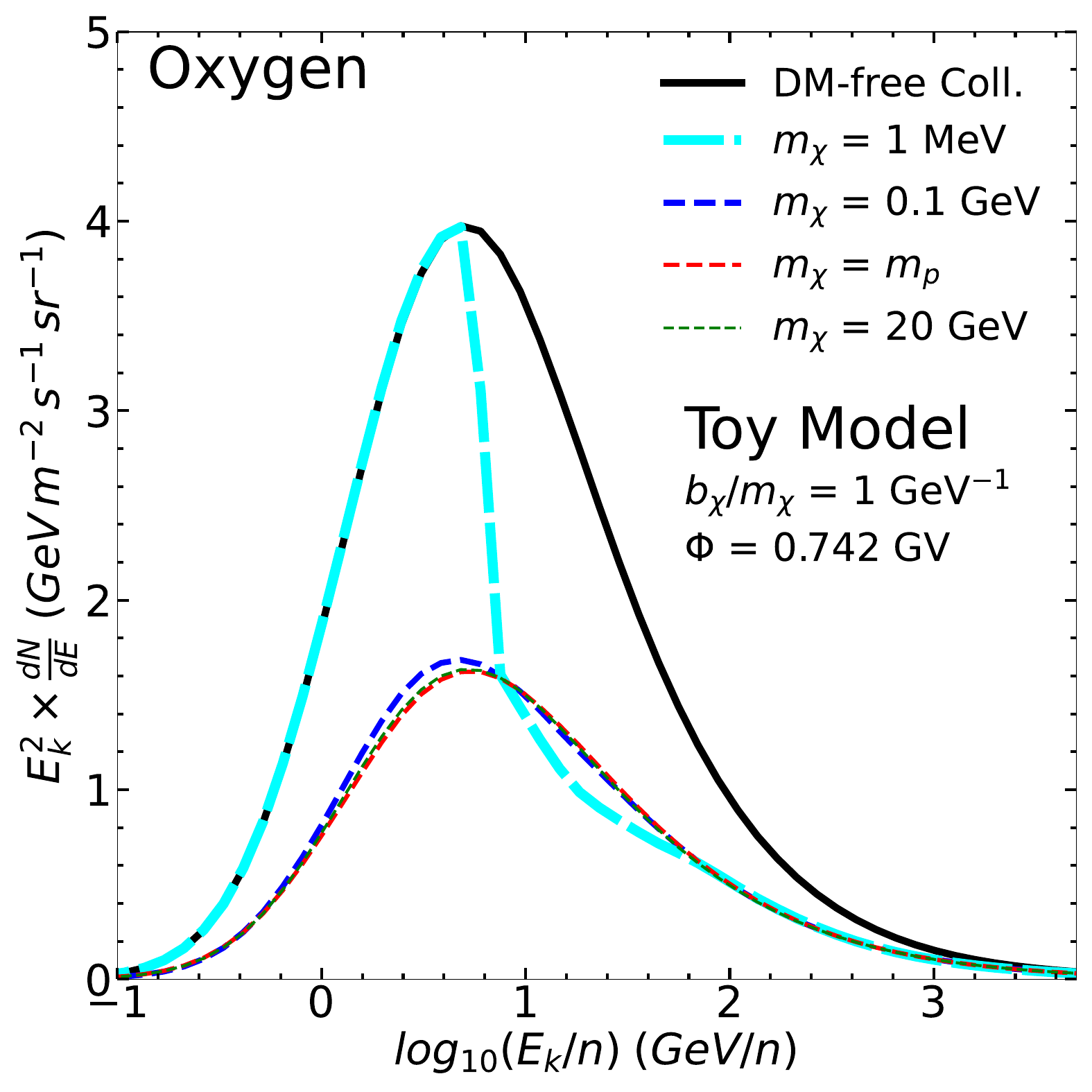}
    \includegraphics[width=0.496\linewidth]{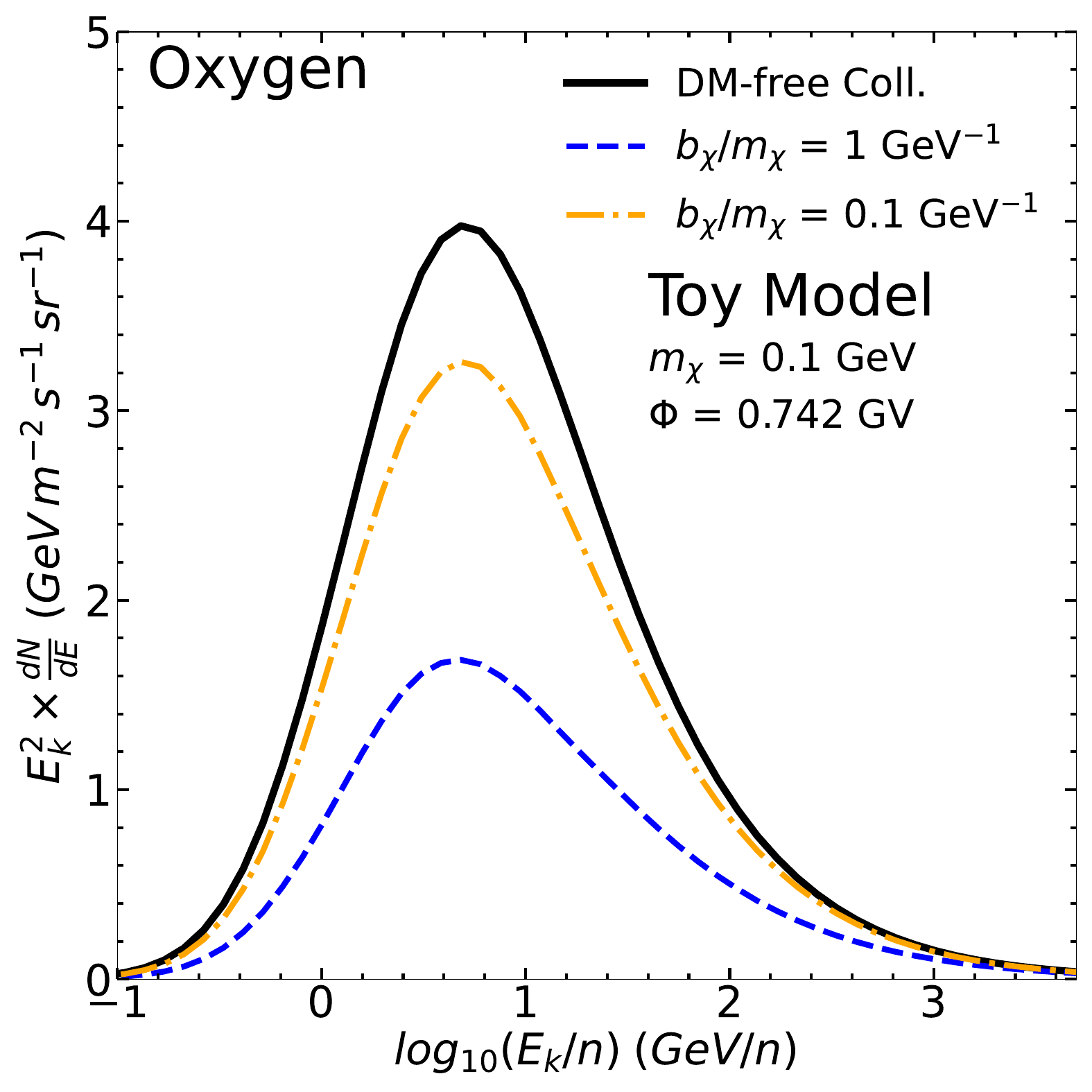}
    \caption{Oxygen energy spectra predicted by the DM-Oxygen toy model in a function of the kinetic energy per nucleon of $^{16}$O, using the propagation parameters from Model A.
    The spectra are shown for four different DM masses, $m_\chi=10^{-3}\gev$ (cyan), $m_\chi=0.1\gev$ (blue), $m_p$ (red), and $20\gev$ (green), 
    with $b_\chi/m_\chi=1\gev^{-1}$ in the left panel. 
    The right panel shows $b_\chi/m_\chi=1\gev^{-1}$ (blue dashed line) and $b_\chi/m_\chi=0.1\gev^{-1}$ (orange dash-dotted line) for comparison.  
    Both lines are based on $m_\chi=0.1\gev$. 
    The black solid lines also give a spectrum for the DM-free scenario. 
    Compared with the DM-free scenario, the fragmentation of Oxygen due to the collision with DM leads to 
    a decrease in the height of the spectrum.}
    \label{fig:toyo}
\end{figure}

\begin{figure}[ht!]
    \centering
    \includegraphics[width=0.496\linewidth]{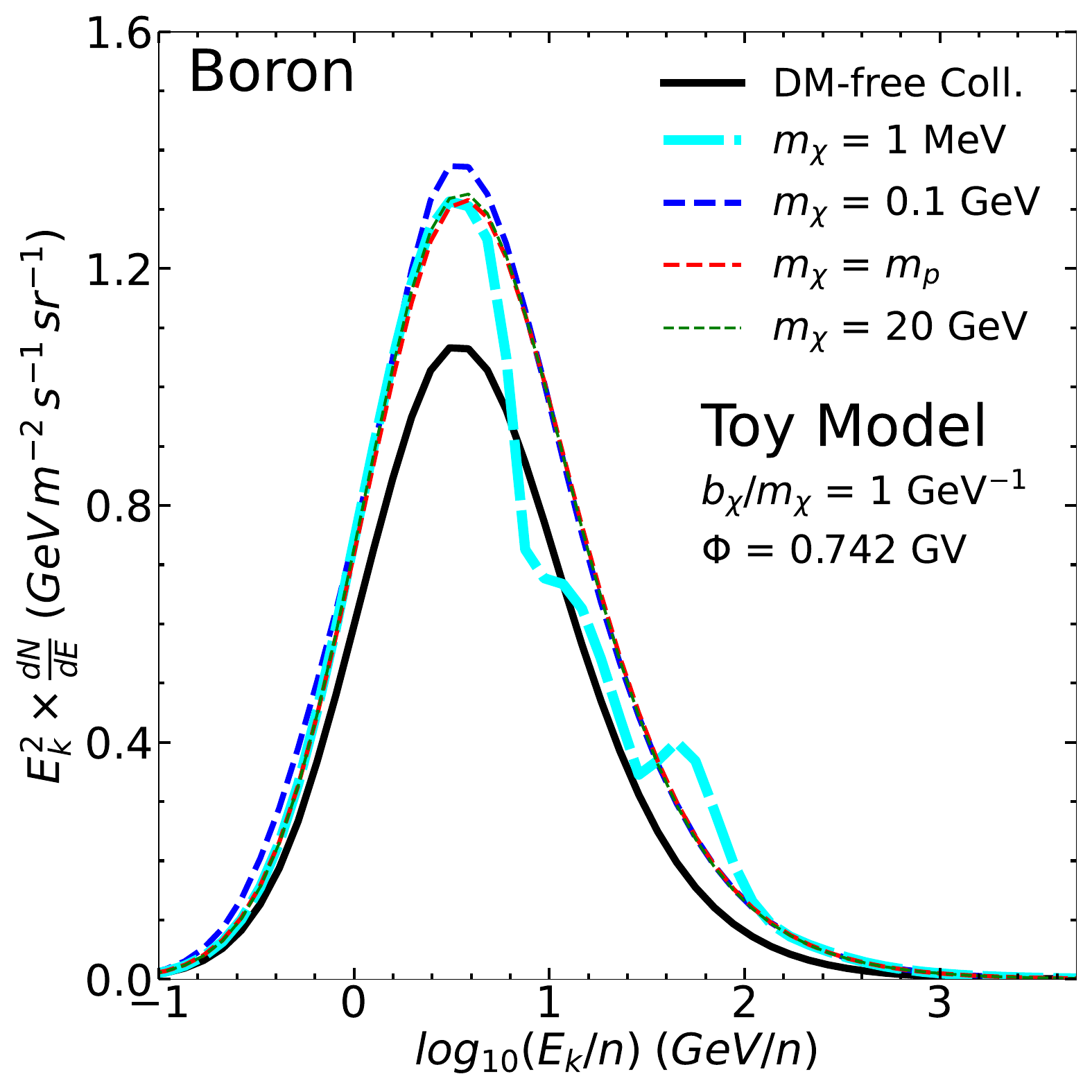}
    \includegraphics[width=0.496\linewidth]{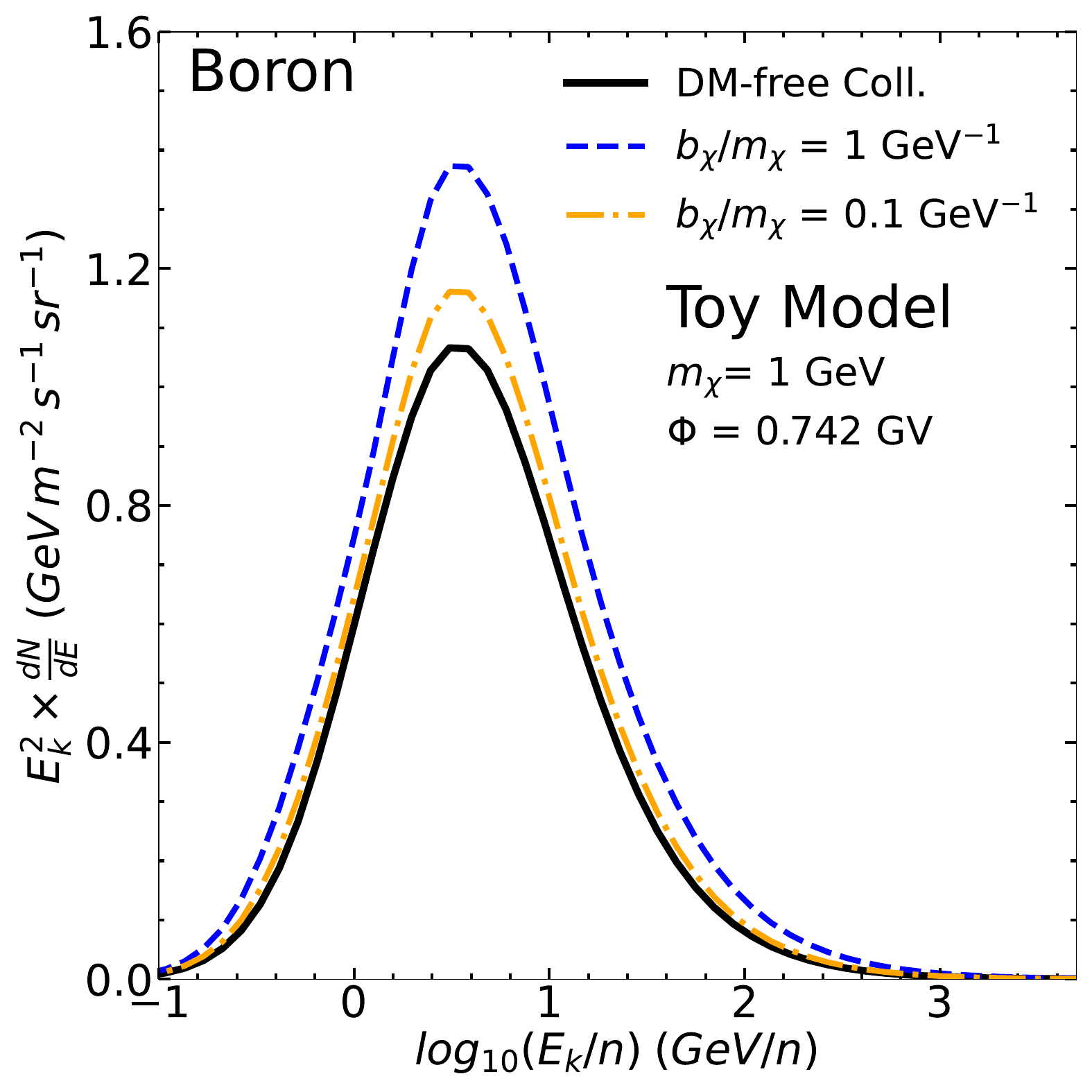}
    \includegraphics[width=0.496\linewidth]{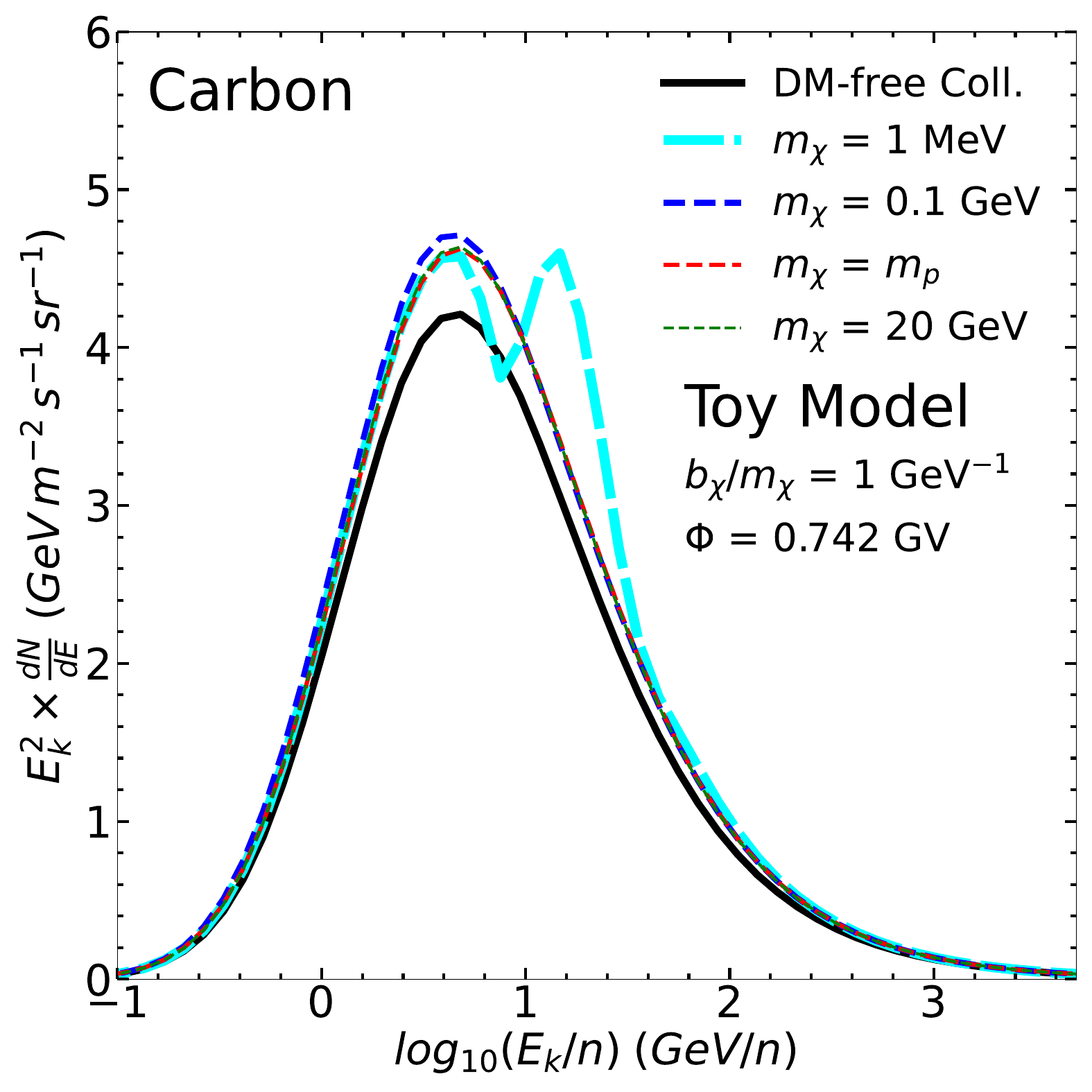}
    \includegraphics[width=0.496\linewidth]{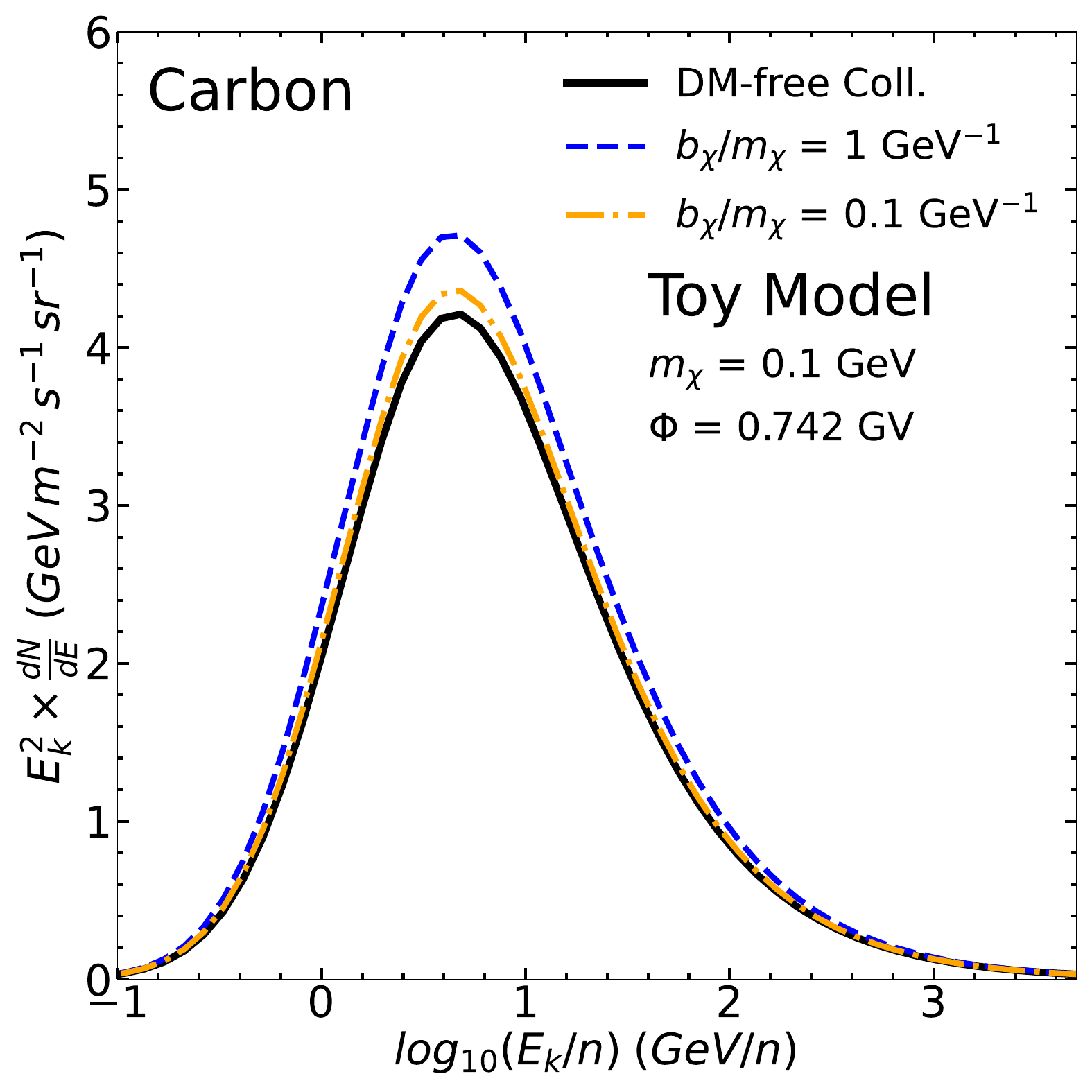}
    \caption{Boron (upper panels) and Carbon (lower panels) energy spectra predicted 
    by the DM-Oxygen toy model in a function of the kinetic energy per nucleon of $^{10}$B, using the propagation parameters from Model A.
    The color scheme is identical to that in Fig.~\ref{fig:toyo}. 
    }
    \label{fig:toyb}
\end{figure}

\begin{figure}[ht!]
    \centering
    \includegraphics[width=0.496\linewidth]{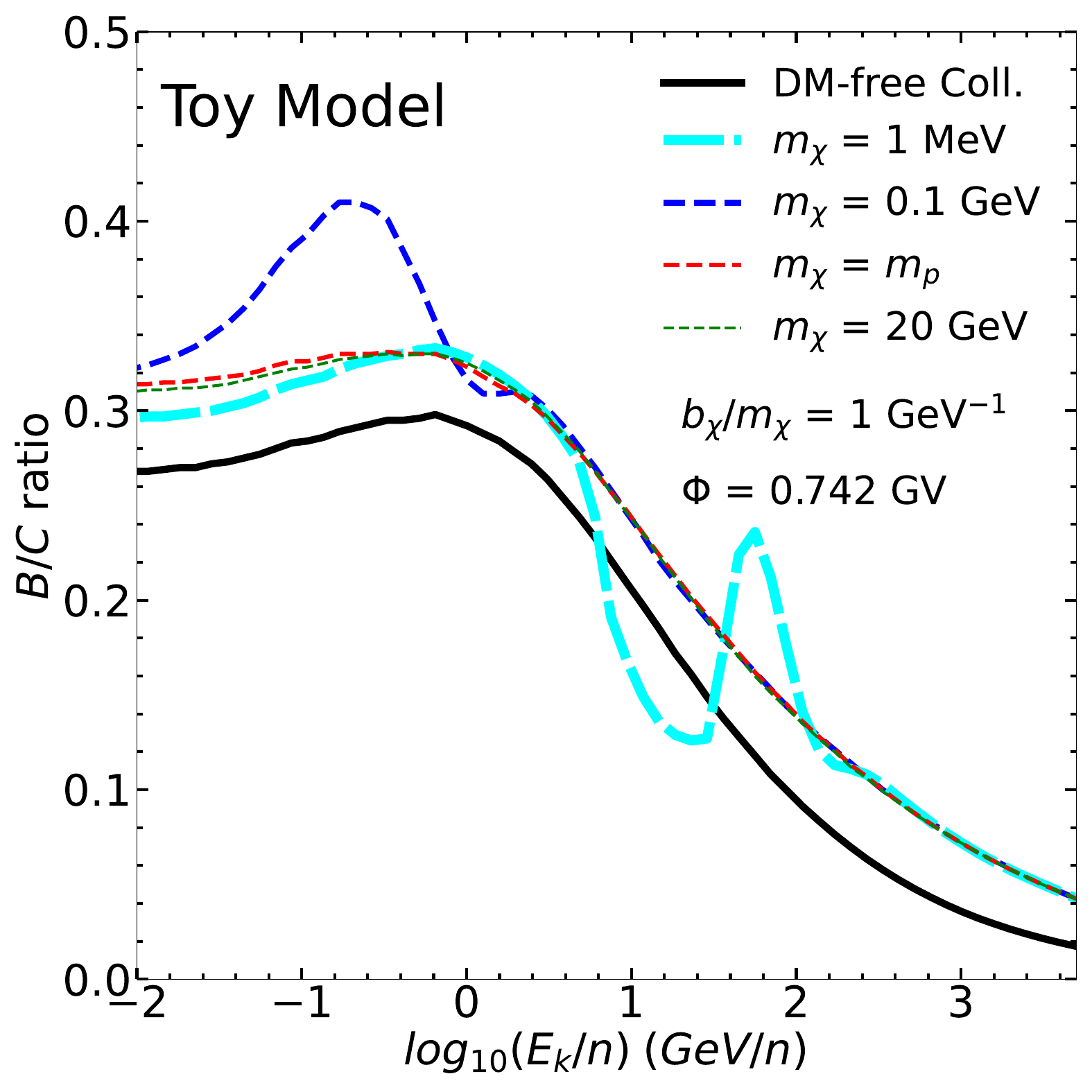}
    \includegraphics[width=0.496\linewidth]{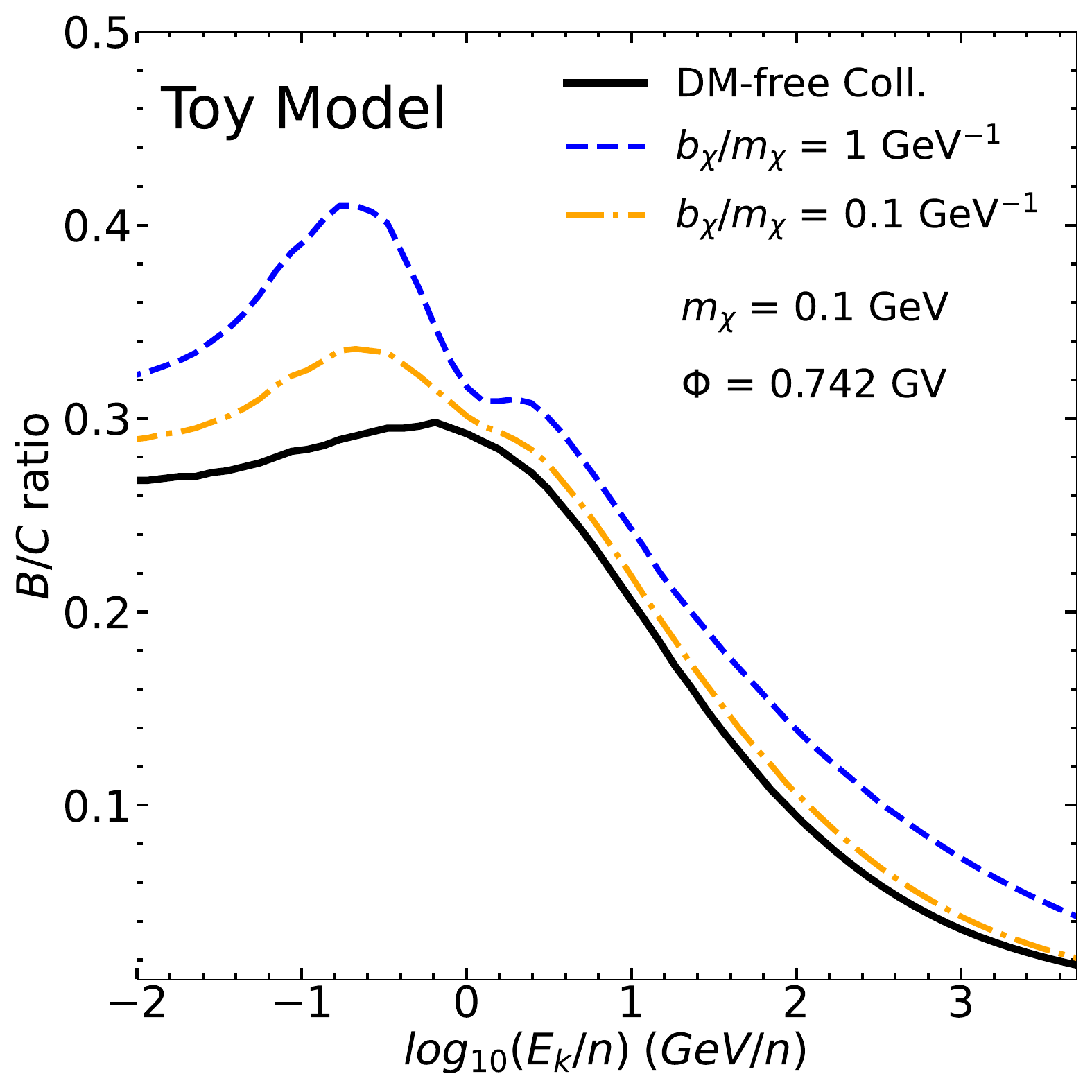}
    \caption{The B/C ratio as a function of the kinetic energy per nucleon, using the propagation parameters from Model A. 
    The color scheme is identical to that in Fig.~\ref{fig:toyo}. }
    \label{fig:toybtc}
\end{figure}

Next, we consider CR energy spectra by varying $b_\chi$ and $m_\chi$ for Oxygen (Fig.~\ref{fig:toyo}), 
Carbon and Boron (Fig.~\ref{fig:toyb}), and the ratio of Boron to Carbon B/C (Fig.~\ref{fig:toybtc}). 
For propagation parameters, we use the values of Model A in Table~\ref{tab:diff-para}.   
As adopted in Fig.~\ref{fig:xsec}, we again show four different DM mass benchmarks, 
$m_\chi=10^{-3}\gev$ (cyan lines), 
$m_\chi=0.1\gev$ (blue lines), 
$m_\chi=m_p$ (red lines) and $m_\chi=20\gev$ (green lines). 
We present $b_\chi=0$ (black solid lines) for a DM-free scenario. 
When comparing the spectra based on different DM masses in these four left panels, we take $b_\chi/m_\chi=1\gev^{-1}$ for a demonstration.  
In these right panels, we plot the spectra produced by CRs collision with $0.1\gev$ DM, 
comparing with $b_\chi/m_\chi=1\gev^{-1}$ and $b_\chi/m_\chi=0.1\gev^{-1}$.

In the left panel of Fig.~\ref{fig:toyo}, 
we can see the spectrum peak position and height slightly depending on 
the value of $m_\chi$. 
Since we only consider the process $^{16}$O+ $\chi \to\cdots$ in this toy model, 
the total Oxygen abundance can be reduced.   
Therefore, a larger fragmentation cross-section, as shown in Fig.~\ref{fig:xsec} can lower the total Oxygen abundance. 
Because of the solar modulation and energy loss resulting from ionization and coulomb scattering within the $E_k\leq 1\gev$ region, 
the spectra of the three heavier scenarios differ from the DM-free scenario only by the height.
Because we use $b_\chi/m_\chi=1\gev^{-1}$ here instead of $b_\chi$ and $m_\chi$, 
the heights of spectra are not too much different for various $m_\chi$.
In the case of $m_\chi=0.1\gev$ compared with the case that $m_\chi=m_p$, 
the flux peak shifts to the smaller energy $E_k\simeq 3\gev$. 
For the heaviest mass case $m_\chi=20\gev$, 
it is hard to find the difference to the $m_\chi=m_p$ case in the spectrum shape. 
Regarding the case with the lightest DM mass $m_\chi=10^{-3}\gev$, 
the Oxygen spectrum behaves like the one in a DM-free scenario at $E_k/n<5\gev$, 
but a suppression happens at $E_k/n \geq 5\gev$. 
Referring to the cyan lines in Fig.~\ref{fig:xsec}, the highest peak of $m_\chi=10^{-3}\gev$ in the total fragmentation cross-section shifts to a higher $E_k/n$ region, but that higher energy spectra are not solar-modulated. 
Therefore, we can see a suppression appears at $E_k/n \geq 5\gev$.

In the right panel of Fig.~\ref{fig:toyo}, we show the $^{16}$O spectra by varying the size of $b_\chi/m_\chi$. 
As we expect, a larger $b_\chi/m_\chi$ makes a large portion of $^{16}$O being fragmented. 
However, the DM-Oxygen collision can be polluted by proton-oxygen collisions so that the total decreasing amount of 
$^{16}$O is not linearly proportional to the ratio $b_\chi/m_\chi$ by comparing two cases $b_\chi/m_\chi=1\gev^{-1}$ (blue dashed line) and 
$b_\chi/m_\chi=0.1\gev^{-1}$ (orange dash-dotted line).

In Fig.~\ref{fig:toyb}, we show the Boron and Carbon spectra for varying $m_\chi$ (two left panels) and $b_\chi/m_\chi$ (two right panels). 
Since the Boron and Carbon are mainly produced by Oxygen fragmentation, the abundance of the Boron and Carbon can be enhanced if increasing $b_\chi/m_\chi$. 
Even if using the same $b_\chi/m_\chi$, only the lightest case $m_\chi=10^{-3}\gev$ differs much from the other three masses 
due to the total fragmentation cross-section shifting to a higher $E_k/n$ region.

In Fig.~\ref{fig:toybtc}, we show the B/C ratio as a function of the kinetic energy per nucleon 
by varying $m_\chi$ (left panel) or $b_\chi/m_\chi$ (right panel). 
Interestingly, we notice that the B/C ratio spectrum appears to be more complex, particularly for  
the peaks of $m_\chi=0.1\gev$ and $m_\chi=10^{-3}\gev$ cases (blue and cyan dashed lines). 
Again, we have learned from Fig.~\ref{fig:xsec} that a DM particle with a mass lighter than the proton mass 
can shift the peak of the fragmentation cross-section toward the higher $E_k/n$ region.  
On the other hand, a heavier DM particle ($m_\chi>m_p$) colliding with $^{16}$O can shift 
the fragmentation cross-section peak to a lower $E_k/n$ region so that 
the cross-section at the higher $E_k/n$ region remains only half of the cross-section at the peak. 
Namely, the spectrum distortion for $m_\chi<m_p$ can be easier identified than the $m_\chi>m_p$ case.  
Moreover, we see from Fig.~\ref{fig:toyb} that the distortion in the Carbon spectra can be stronger than the distortion in the Born spectra. 
Hence, the distortion in the B/C ratio spectra for $m_\chi<m_p$ case can be even more measurable than $m_\chi>m_p$ if $b_\chi/m_\chi$ is large. 
When decreasing $b_\chi/m_\chi$ as shown in the right panel of Fig.~\ref{fig:toybtc}, 
the distortion can be alleviated as expected.

\subsection{The CR spectra including full DM-CRs collisions}
\label{sec:gen}

\begin{figure}[ht!]
    \centering
    \includegraphics[width=0.496\linewidth]{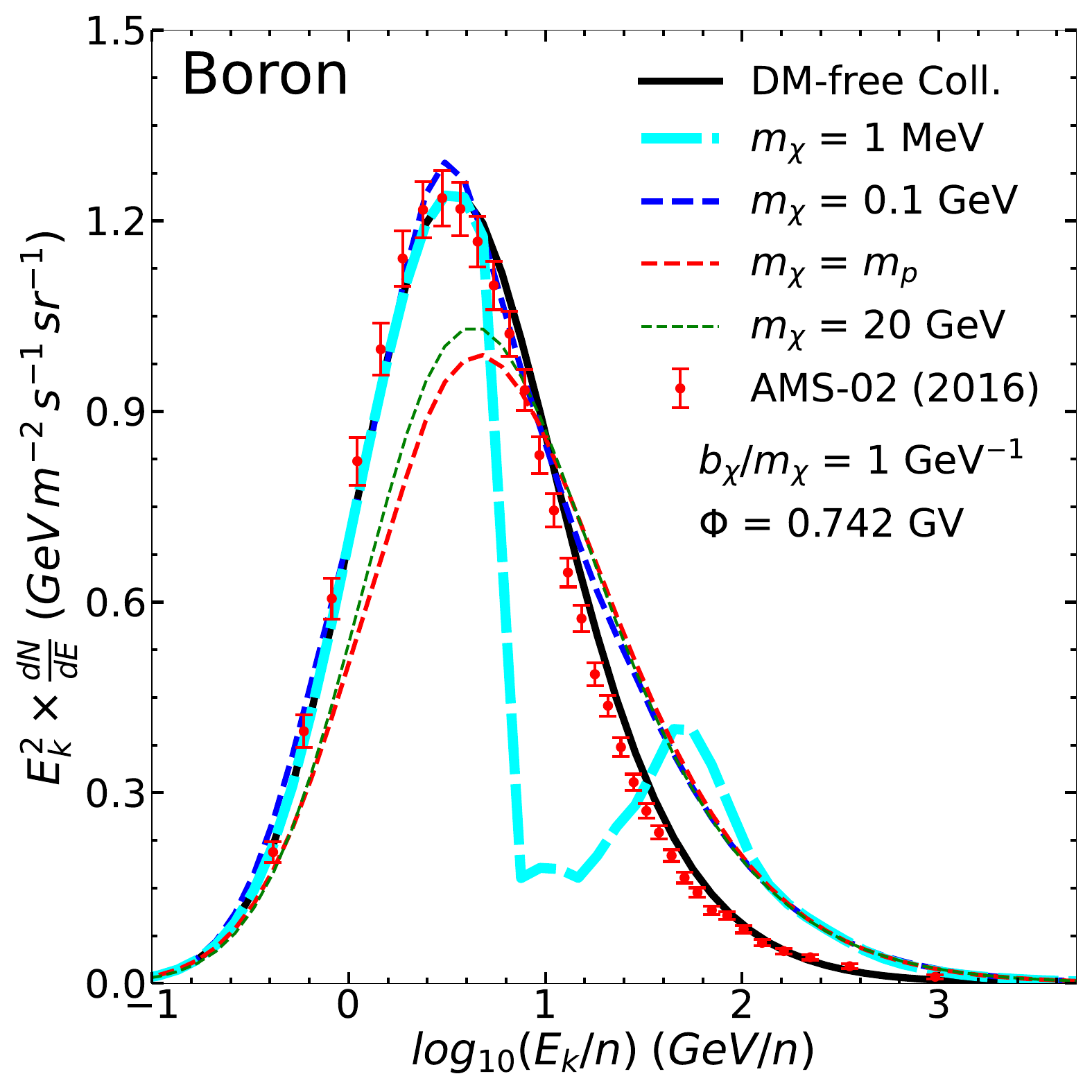}
    \includegraphics[width=0.496\linewidth]{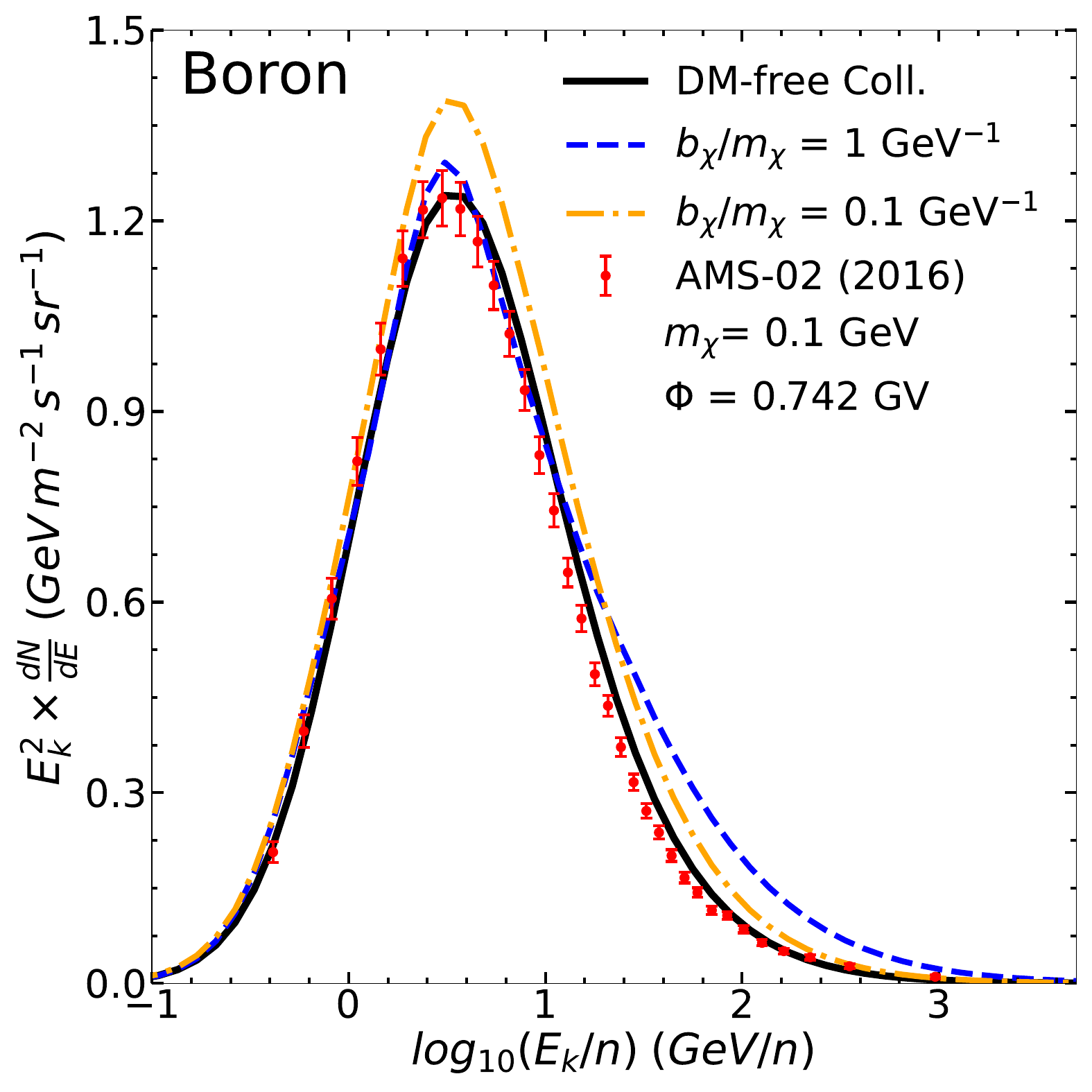}
    \includegraphics[width=0.496\linewidth]{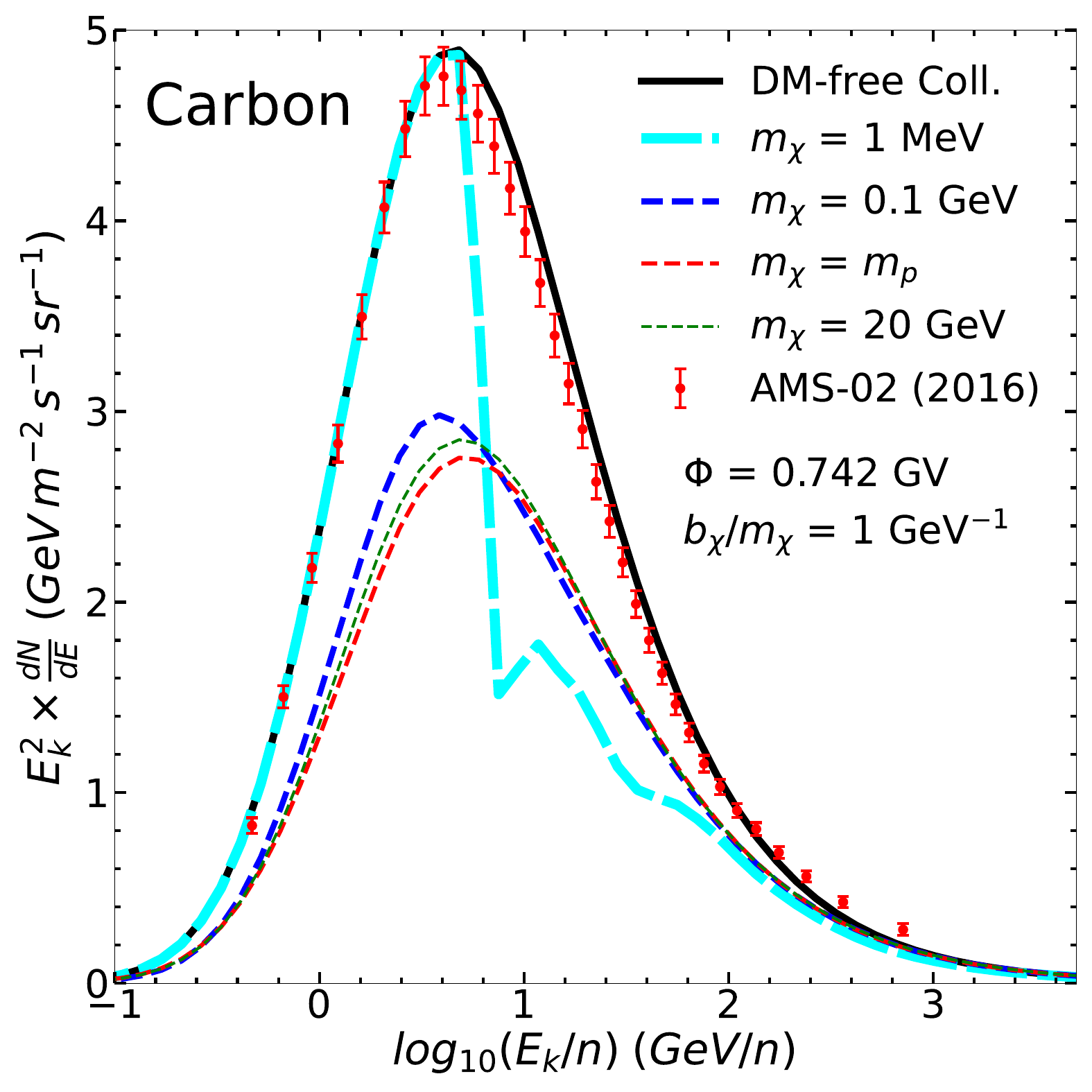}
    \includegraphics[width=0.496\linewidth]{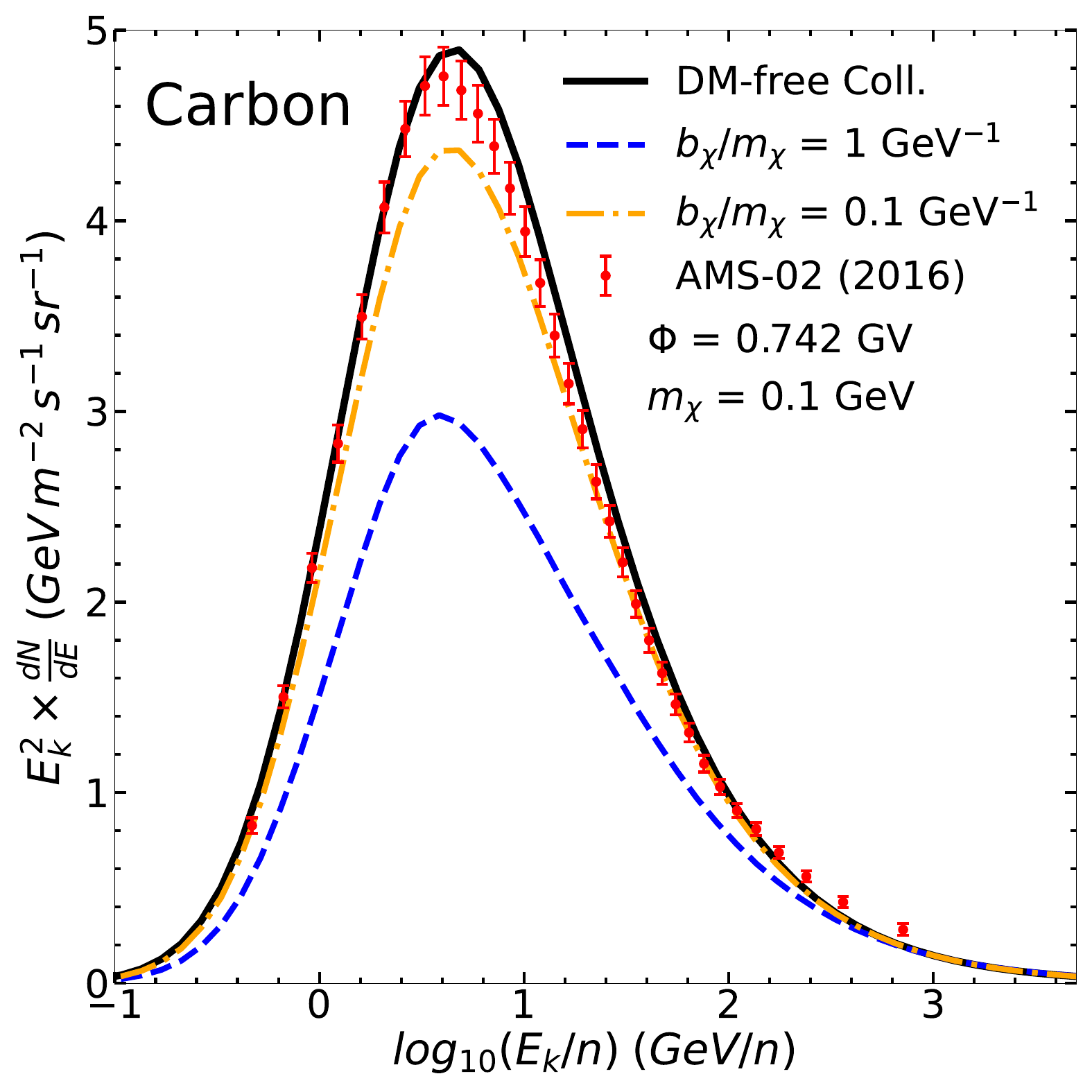}    
    \caption{The energy spectra for Boron (two upper panels) and Carbon (two lower panels) with the propagation parameters given by Model A. 
    The full interactions between DM and CRs are taken into account. 
    The color scheme is the same as Fig.~\ref{fig:toyo}, while  
    the AMS-02 data are presented as red error bars. 
Two left panels show the spectra change with respect to $m_\chi$, but 
the two right panels compare two different $b_\chi/m_\chi$ by fixing $m_\chi$. }
    \label{fig:Carbonreal}
\end{figure}

In this subsection, we move to a general case that DM can collide with all the CR particles, 
from hydrogen ($Z=1$) to nickel ($Z=28$).  
Apparently, the cascade and fragmentation processes after DM-CR collisions become complicated and hard to backtrack precisely. 
The origins of the spectrum growth or reduction can be already washed out. 
Fortunately, the peak of the CR spectra still has some similarities with 
the toy model (DM-Oxygen collision only) demonstrated in Sec.~\ref{sec:toy}.  
Hence, we will only focus on the spectra of Carbon, Boron, and the B/C ratio.

Unlike the toy model presented in Fig.~\ref{fig:toyb},  
Carbon spectra in two lower panels of Fig.~\ref{fig:Carbonreal} are smaller than the spectra without DM-CRs collisions, 
even if using $b_\chi/m_\chi=0.1\gev^{-1}$ (the orange dash-dotted line in the lower right panel).  
We find that the DM-Carbon fragmentation cross-section 
is almost twice higher than the production cross-section of the $\chi+O\to C$ process, 
especially for the region with $E_k/n$ higher than peak energy.  
On the other hand, Carbon and Oxygen abundance in the cosmic ray are comparable. 
Hence, the DM-induced spallation rate is faster for the Carbon case than the DM-induced production rate.  
In the Boron case, the CR Boron abundance is lower than the CR Carbon abundance, but 
the cross-sections of Boron produced by Oxygen and Carbon are higher than the Boron fragmentation cross-section. 
Once a smaller Boron fragmentation cross-section is applied, like the orange dash-dotted line in the upper right panel, 
the Boron production is sufficient to enhance the Boron spectrum to be higher than the DM-free scenario.

\begin{figure}[ht!]
    \centering
    \includegraphics[width=0.496\linewidth]{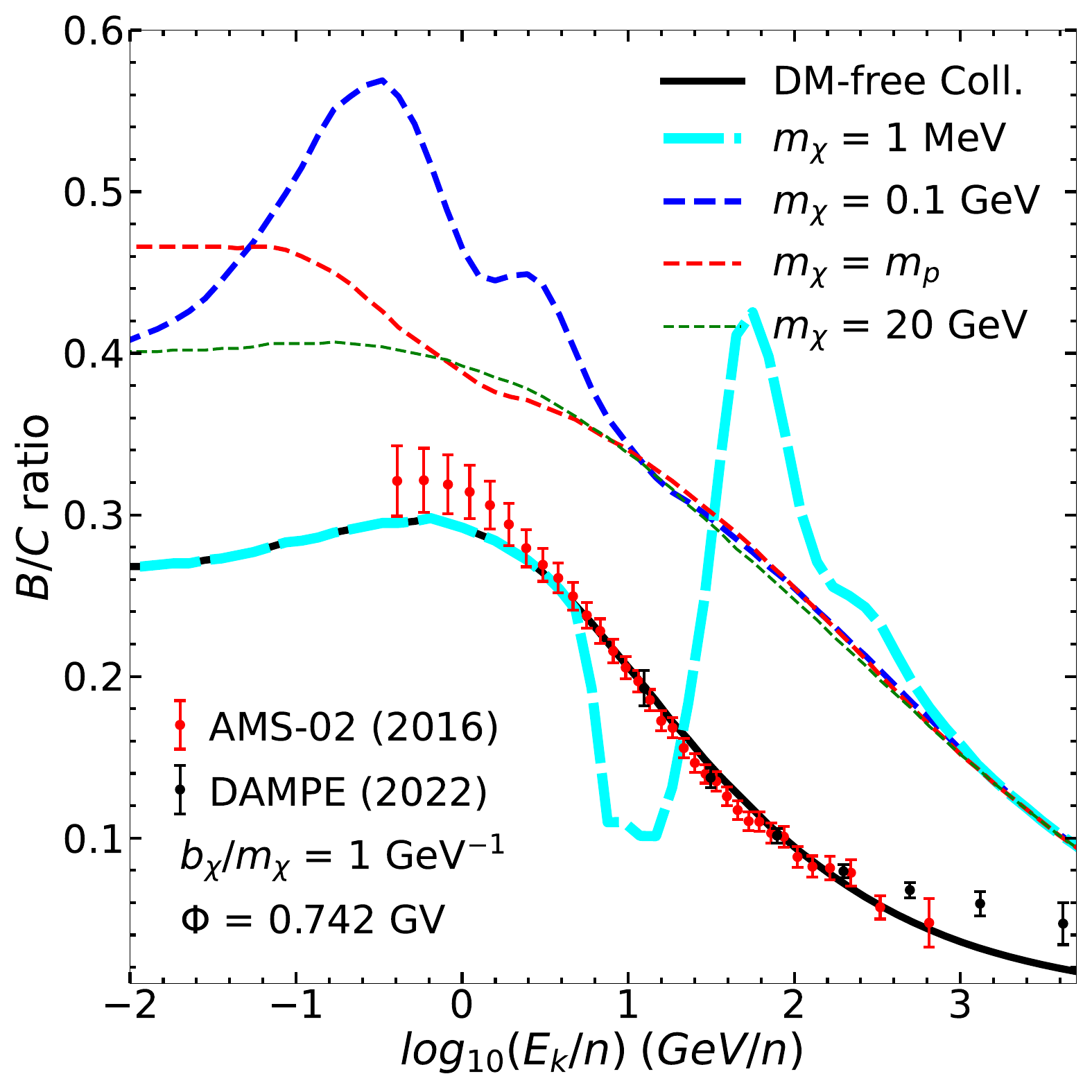}
    \includegraphics[width=0.496\linewidth]{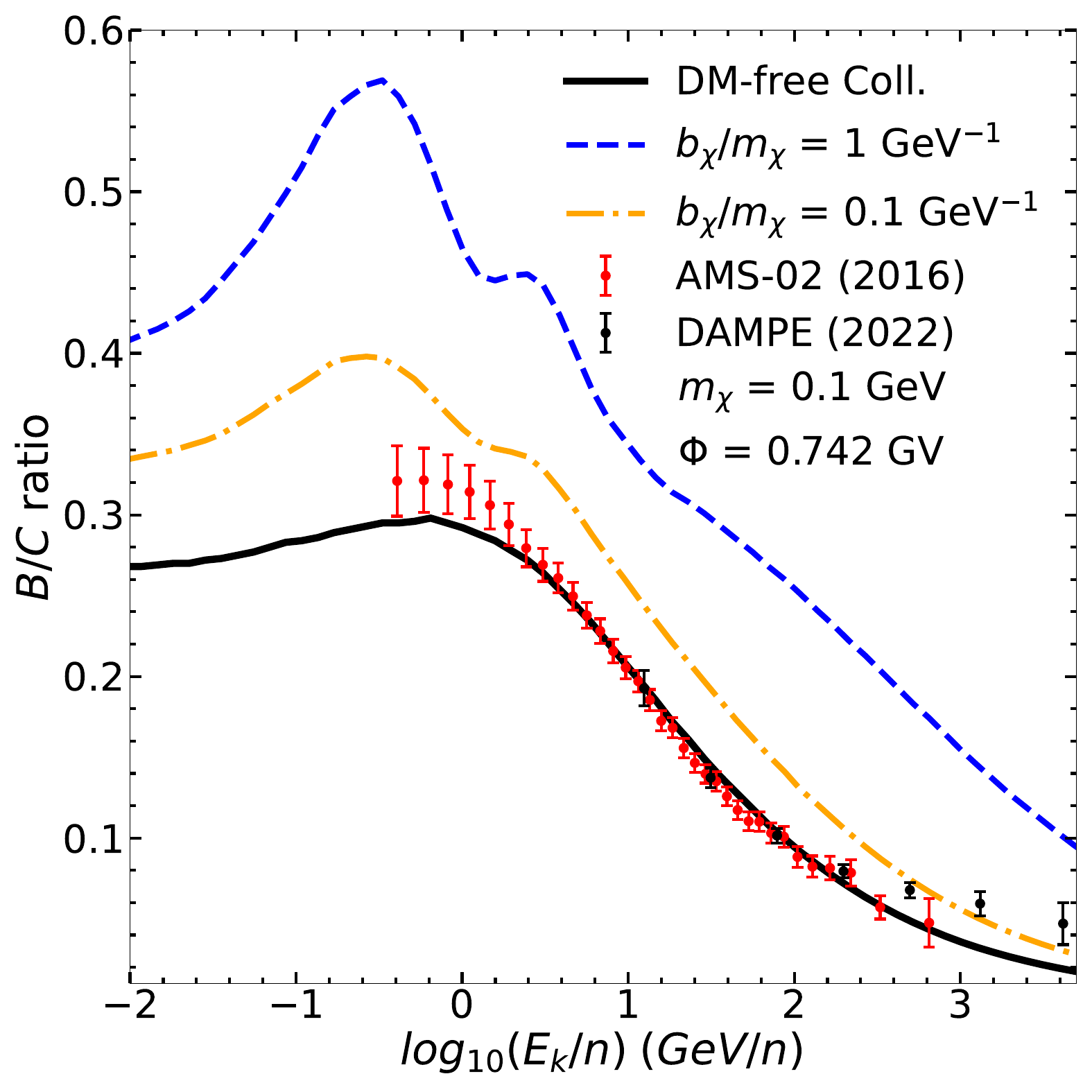}
    \caption{The B/C ratio similar to Fig.~\ref{fig:toybtc} but for DM-CRs collisions fully implemented. 
    The experimental data from AMS-02 and DAMPE are presented as red and black error bars.} 
    \label{fig:realbtc}
\end{figure}

In Fig.~\ref{fig:realbtc}, we plot the B/C ratio by including all DM-CRs collisions. 
In the left panel, we again compare four benchmark DM masses. 
Overall speaking, the spectrum shapes are similar to Fig.~\ref{fig:toybtc}, 
but the enhancement for $m_\chi=m_p$ case at the $E_k/n<1\gev$ region is due to the cascade contributions from 
other heavier element fragmentation. 
Again, the peak feature of the green line ($m_\chi=20\gev$) is smeared out because of the solar modulation. 
However, the DM-induced peaks still appear in the light DM cases.  
Compared with Fig.~\ref{fig:toybtc}, the B/C ratios in the right panel of Fig.~\ref{fig:realbtc} are generally greater, owing to the fact that the Boron spectrum with the full DM-CRs collisions is higher than the Carbon one as shown in Fig.~\ref{fig:Carbonreal}.   
Even if taking $b_\chi/m_\chi=0.1\gev^{-1}$, we can still clearly distinguish the DM-induced B/C ratio from the one without DM contribution.

\subsection{Impacts of $D_0$ and $Z_h$ on the B/C ratio}
\label{sec:astro_unc}

\begin{figure}[ht!]
    \centering
    \includegraphics[width=0.496\linewidth]{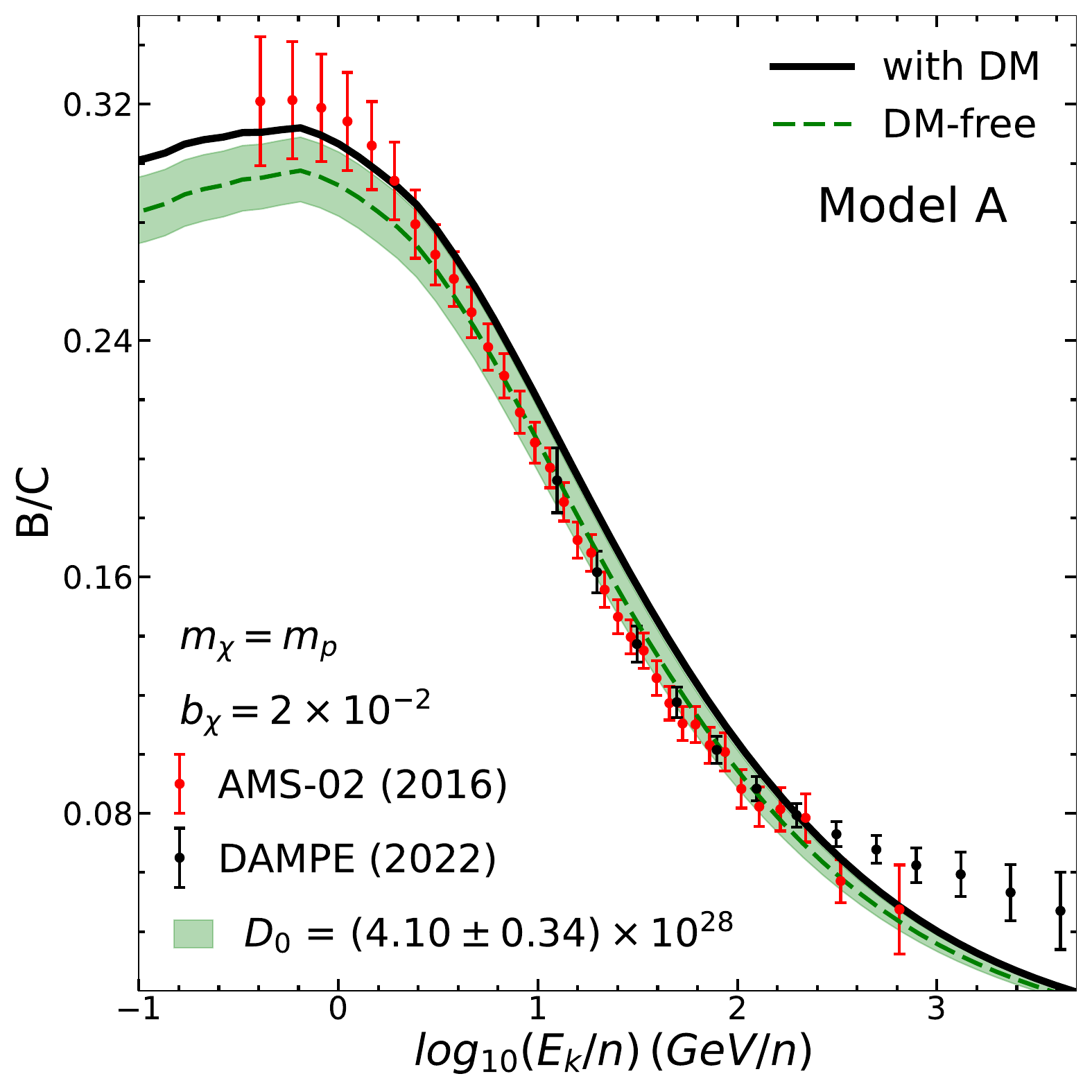}
    \includegraphics[width=0.496\linewidth]{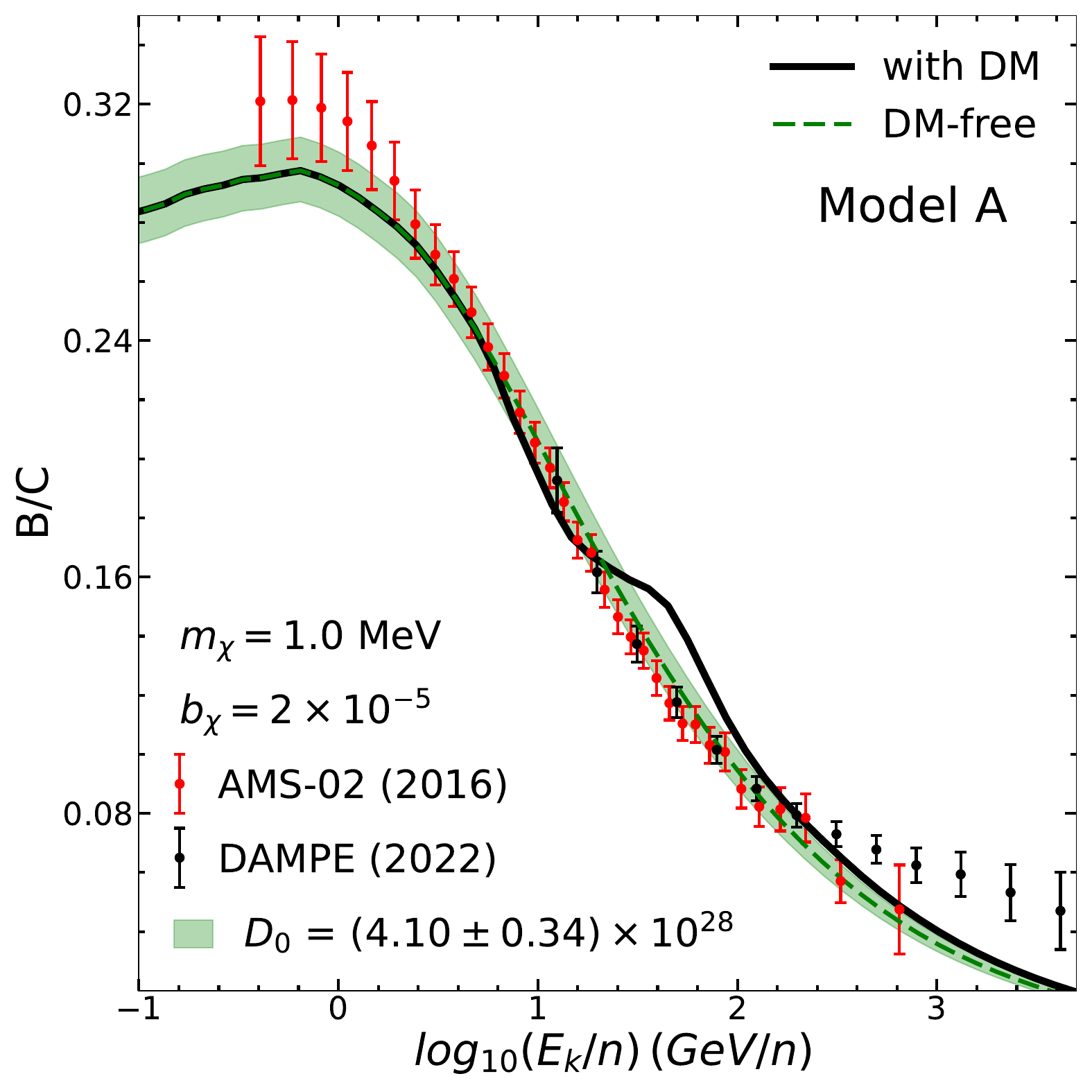}
    \includegraphics[width=0.496\linewidth]{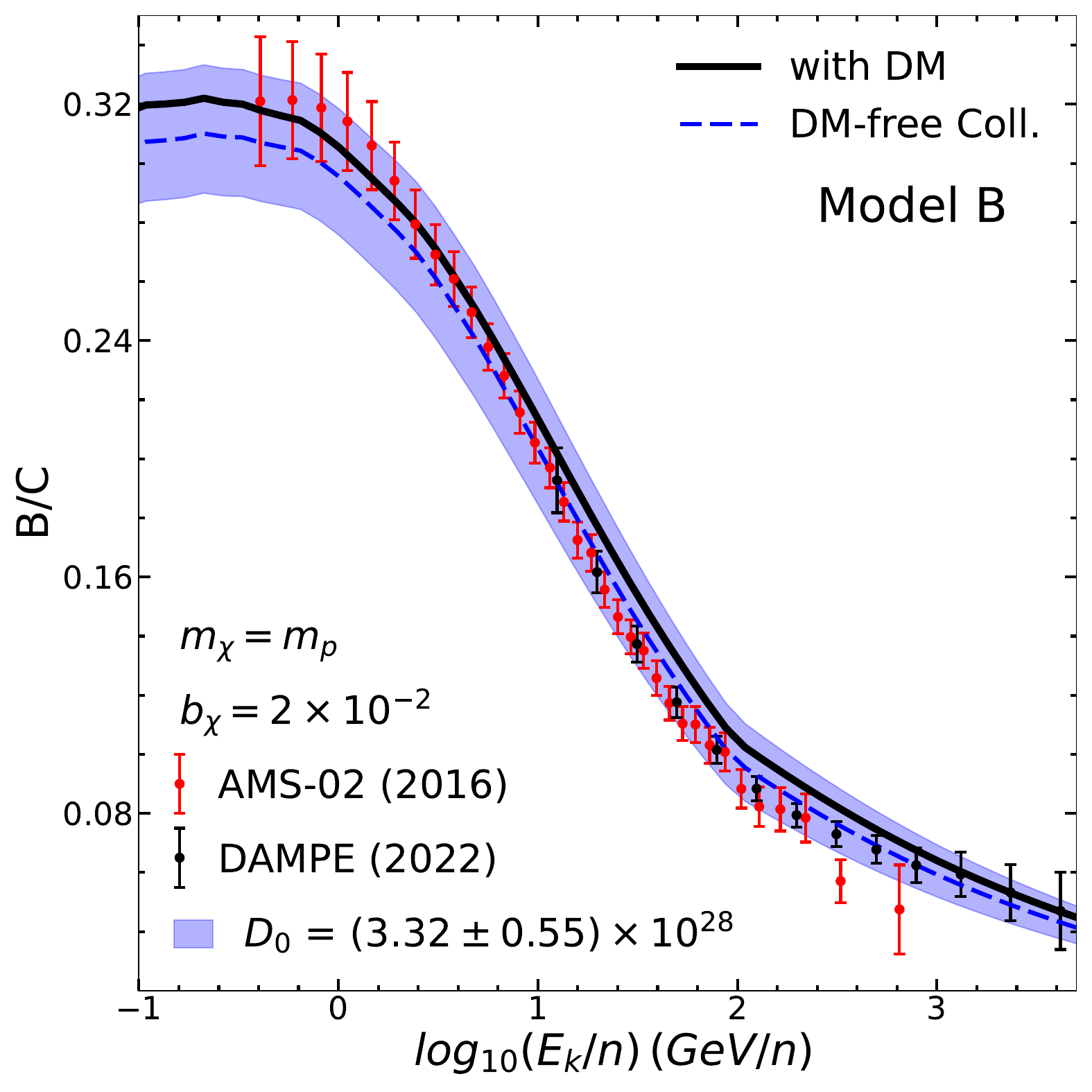}
    \includegraphics[width=0.496\linewidth]{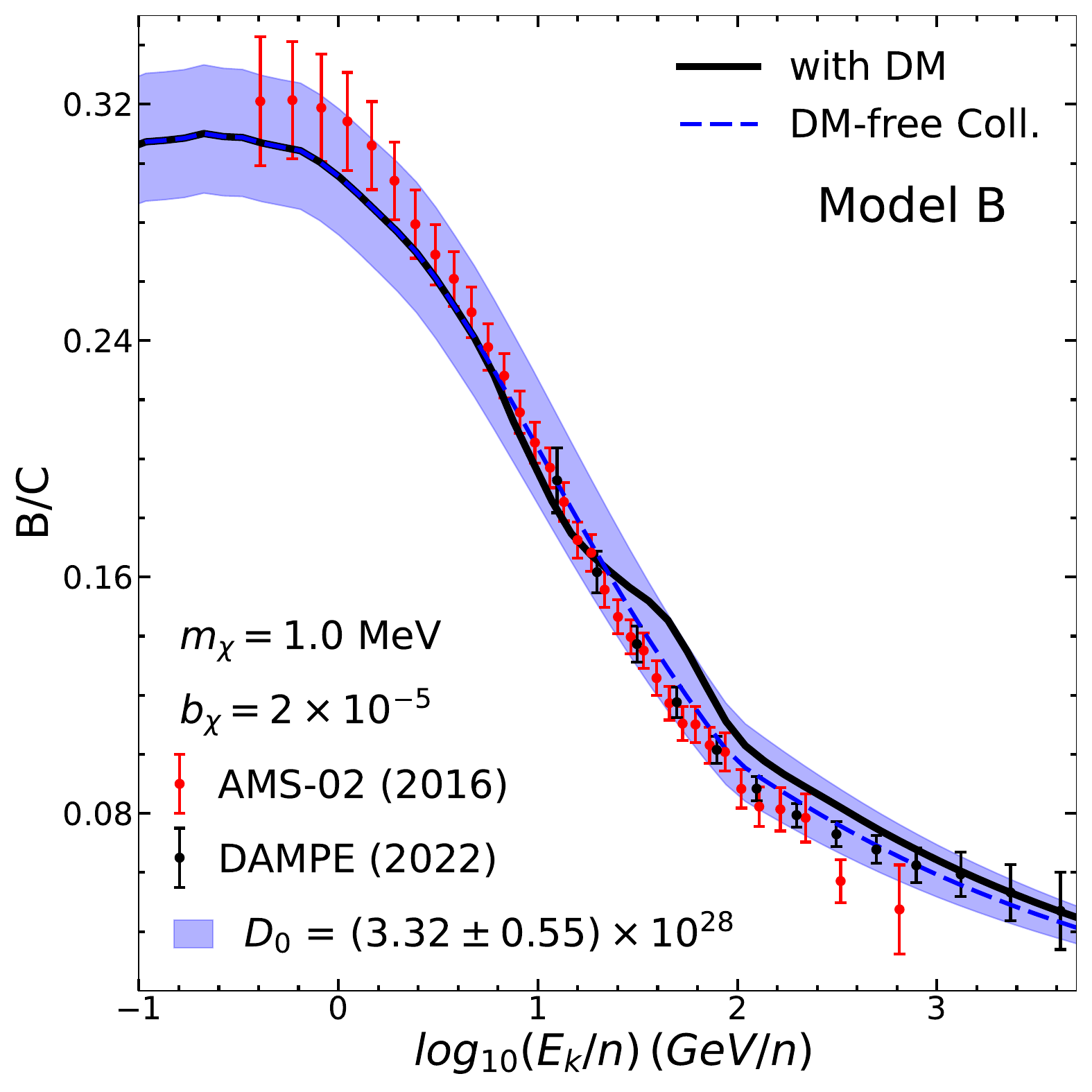}
    \caption{The B/C ratio spectra for two different masses, $m_\chi=m_p$ (left panels) and $m_\chi=1\mev$ (right panels). 
The black solid lines depict the spectra with $b_\chi/m_\chi=0.02$ using Model A (upper panels) and Model B (lower panels), while the dashed lines show spectra for the DM-free scenarios using Model A (green) and B (blue). The green and blue-shaded regions represent the DM-free scenarios by varying $D_0$ within the $1\sigma$ region from Model A and Model B, while the red and black error bars denote the $1\sigma$ of AMS-02 measurements~\cite{AMS:2016brs} and the DAMPE measurements~\cite{2210.08833}.} 

    \label{fig:best_ben_heavy}
\end{figure}

The undetermined propagation parameters, such as $D_0$ and $Z_h$ may introduce additional uncertainties.  
Thus, the DM-induced distortion of the B/C ratio becomes hard to distinguish from the propagation uncertainties. 
For simplicity, we analyze the impacts using the one sigma regions of the diffusion coefficient ($D_0/(10^{28}~{\rm cm}^2 {\rm s}^{-1})=4.10\pm 0.34$) and the semi-height of the diffusive zone ($Z_h/{\rm kpc}=4.93\pm 0.53$) from Model A~\cite{1701.06149,1810.03141}. 
For Model B, the values are $D_0/(10^{28}~{\rm cm}^2 {\rm s}^{-1})=3.32\pm 0.55$ and $Z_h/{\rm kpc}=3.61\pm 0.69$~\cite{2210.09205}, 
all consistent with the fitted value of the radioactive cosmic ray and radio data ($Z_h=4.1^{+1.3}_{-0.8}$) within a confidence level of approximately 1$\sigma$~\cite{2004.00441}.
However, Ref.~\cite{1980Ap&SS..68..295G} shows that the CR propagation is sensitive to $D_0$ and $Z_h$, but two parameters degenerate.
Instead of considering both $D_0$ and $Z_h$, we only demonstrate the impacts of $D_0$ variation on the DM-CRs collision-induced B/C ratio in this subsection.

From Fig.~\ref{fig:realbtc}, we find no sharp peak in the B/C ratio due to DM collisions 
for $m_\chi>m_p$, and their B/C ratio spectra can be only overall scaled.  
Hence, we only take $m_\chi\le m_p$ as examples to investigate two different DM masses ($m_\chi=m_p$ and $1\mev$).

In Figure \ref{fig:best_ben_heavy}, we display the B/C ratio spectra for two different DM masses. The black solid lines depict the B/C spectra for scenarios with $b_\chi/m_\chi\approx0.02$, using propagation parameters from Model A (upper panels) and Model B (lower panels). The green dashed lines (Model A) and blue dashed lines (Model B) represent the B/C spectra of DM-free scenarios. Additionally, the green and blue bands show the $1\sigma$ region of the $D_0$ value. 
It is worth noting that similar B/C ratio spectra are observed in Models A and B when $m_\chi$ is fixed. 

When $m_\chi=m_p$, there is no sharp peak observed in the B/C ratio spectra. 
In the case of a lighter DM mass ($m_\chi=1\mev$), a gap appears at around $E_k/n\sim200\gev$, 
followed by a peak at $E_k/n>400\gev$, consistent with Fig.~\ref{fig:realbtc}.
The upper edge of the blue band corresponds to $D_0=2.77\times 10^{28}$~cm$^2$s$^{-1}$ ($-1\sigma$ region), 
while the lower edge corresponds to $D_0=3.87\times 10^{28}$~cm$^2$s$^{-1}$ ($+1\sigma$ region). 
For the green band, the upper and lower edges correspond to $D_0=3.76\times 10^{28}$~cm$^2$s$^{-1}$ and $4.44\times 10^{28}$~cm$^2$s$^{-1}$.
We plot the red and black error bars as the $1\sigma$ region for AMS-02~\cite{AMS:2016brs} and DAMPE~\cite{2210.08833} measurements.  
Regardless of the propagation models, changing $D_0$ results in a similar variation in the B/C ratio spectrum.
\subsection{Constraints on DM parameters}
\label{sec:upperlimits}

We now proceed to estimate the $95\%$ upper limits of the parameter $b_\chi$ by using the B/C data from AMS-02 and DAMPE.   
Our statistic strength $\delta\chi^2$ with the given parameters \{$D_0, Z_h, m_\chi, b_\chi$\} is defined as  
\begin{equation}
\delta\chi^2(D_0, Z_h, m_\chi, b_\chi) =\chi_{\rm DM}^2(D_0, Z_h, m_\chi, b_\chi)-
\chi_{\rm Bkg}^2(D_0, Z_h), 
\label{eq:findBF}
\end{equation}
where DM-induced $\chi_{\rm DM}^2(D_0, Z_h, m_\chi, b_\chi)$ and 
DM-free $\chi_{\rm Bkg}^2(D_0, Z_h)$ are calculated with the expression given in 
Eq.~\ref{eq:chisq}. 
For one-sided $95\%$ upper limit of DM-CR interaction strength $b_\chi^{\rm 95\%}$, 
we require $\delta\chi^2(D_0, Z_h, m_\chi, b_\chi^{\rm 95\%})=2.71$ by fixing $D_0$, $Z_h$, and $m_\chi$.

\begin{figure}[ht!]
    \centering
    \includegraphics[width=0.496\linewidth]{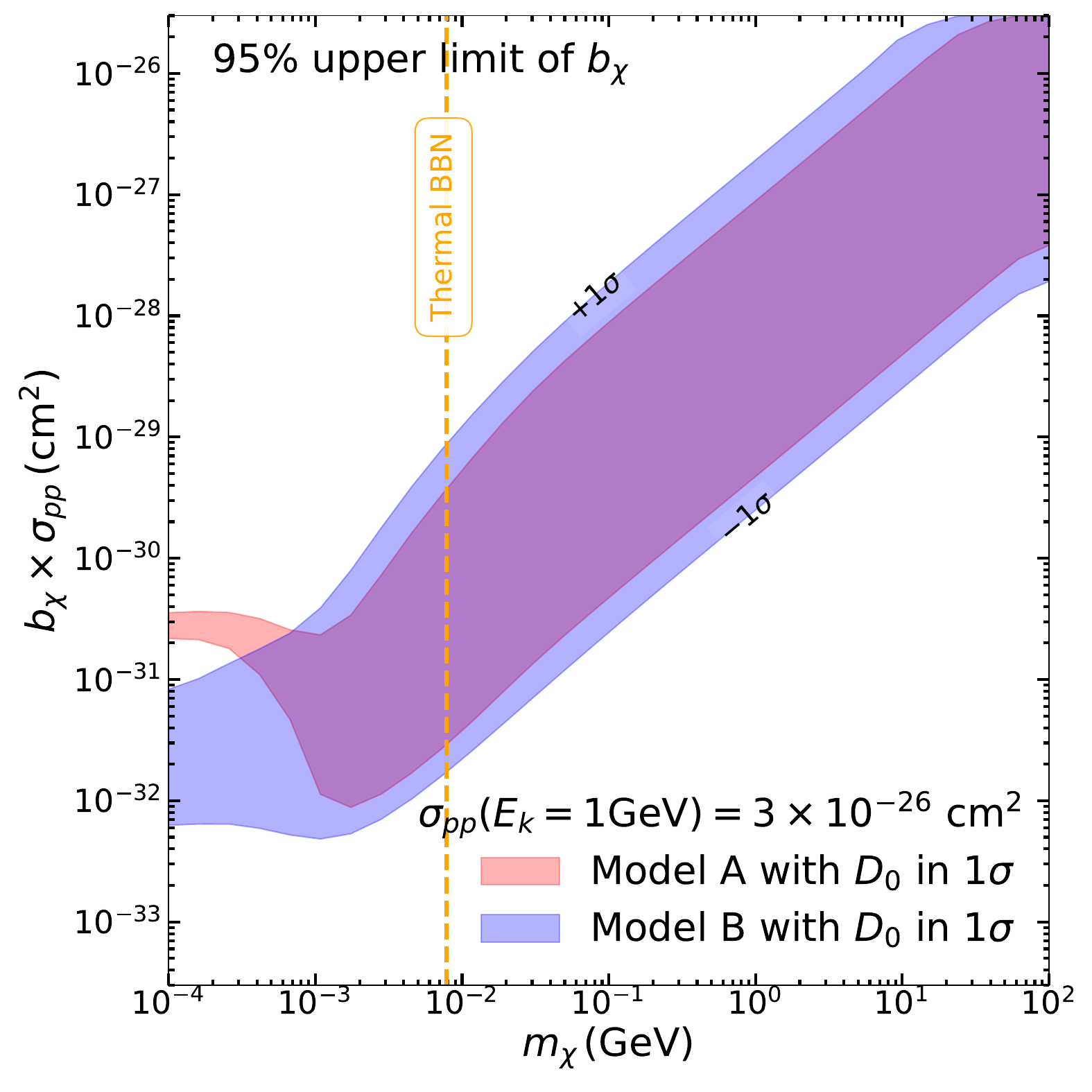}
    \includegraphics[width=0.496\linewidth]{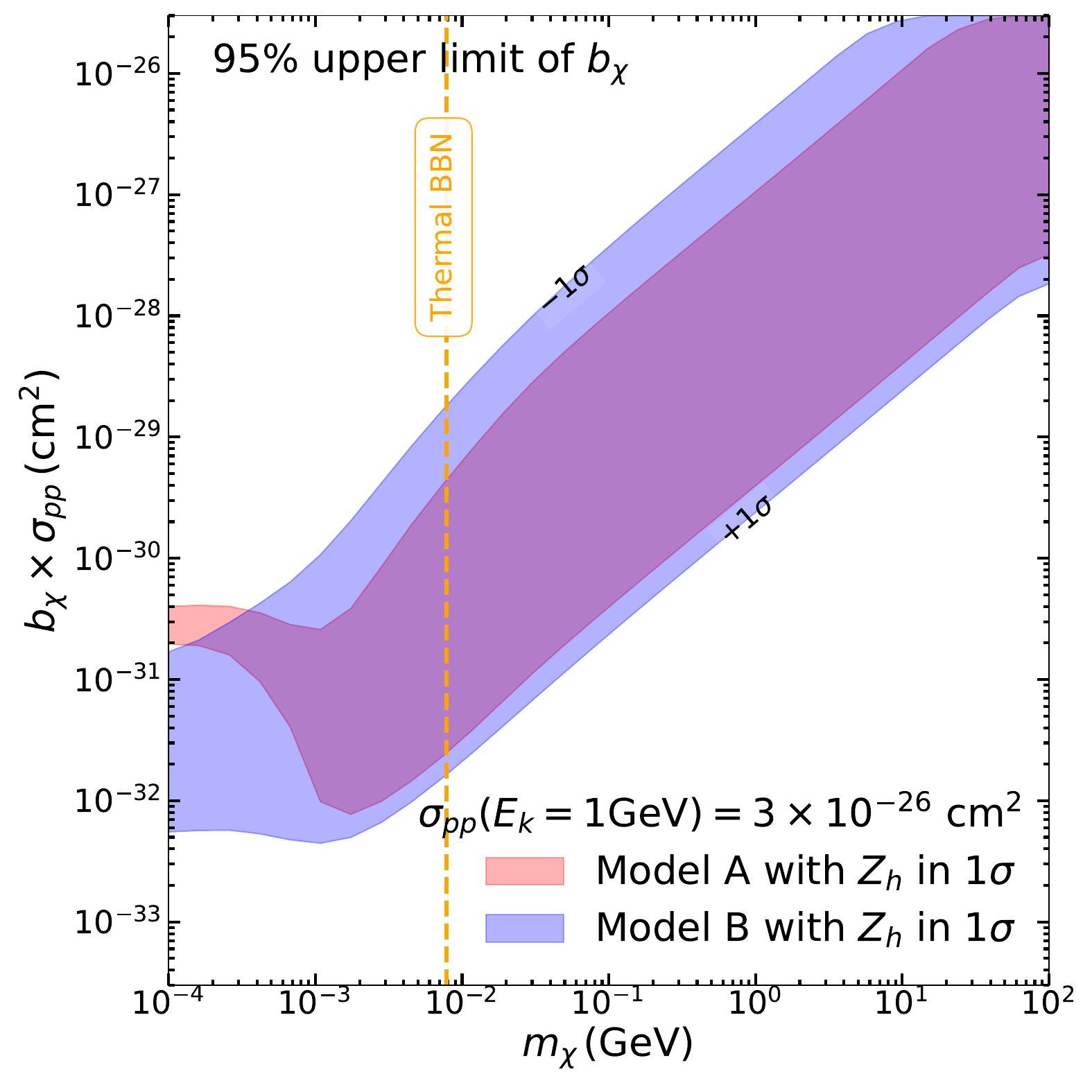}
    \caption{
    The $95\%$ upper limit of $b_\chi$. 
    Instead of plotting $b_\chi$ in the $y$-axis, 
    we show $b_\chi\times\sigma_{pp}$ to enable to cross-compare with proton-proton collision. 
    The value $\sigma_{pp}=3\times 10^{-26}$~cm$^2$ is the proton-proton collision cross-section with $E_k=1\gev$. 
    The upper and lower edges of the shaded regions represent the $1\sigma$ bounds for $D_0$ (left panel) and $Z_h$ (right panel). 
    The blue bands (Model A) present a $1\sigma$ range of $D_0/(10^{28}~{\rm cm}^2 {\rm s}^{-1})=4.10\pm 0.34$ and $Z_h/{\rm kpc}=4.93\pm 0.53$, whereas the red bands (Model B) are defined by $D_0/(10^{28}~{\rm cm}^2 {\rm s}^{-1})=3.32\pm 0.55$ and $Z_h/{\rm kpc}=3.61\pm 0.69$. 
    For reference, the thermal BBN bound $m_\chi \leq 7.8\mev$ for a Dirac fermion DM is represented by an orange dashed line~\cite{1908.00007}. 
    }  
    \label{fig:exclusion}
\end{figure}

In Fig.~\ref{fig:exclusion}, we present the 95\% upper limit for the parameter $b_\chi$, along with the $1\sigma$ uncertainty bands for parameters $D_0$ (left panel) and $Z_h$ (right panel). The uncertainty bands derived from Model A are depicted in red, while those associated with Model B are shown in blue. In the context of a thermally produced DM scenario that involves a Dirac fermion DM and a complex scalar mediator, constraints from Big Bang Nucleosynthesis (BBN) on $\Delta N_{\rm eff}$ establish a lower bound, suggesting that $m_\chi \leq 7.8\ \mathrm{MeV}$, as illustrated by the orange dashed lines~\cite{1908.00007}. For further reference, Table~\ref{tab:diff-para} lists the central values of the propagation parameters. Additionally, compared to Model A, Model B exhibits broader $1\sigma$ regions for both $D_0$ and $Z_h$, leading to the blue bands being wider than the red bands. 
We find three different behaviors in the $95\%$ upper limits: 
\begin{enumerate}
    \item in the first mass region ($m_\chi>0.1\gev$),  this region exhibits a trivial behavior because the B/C ratios are approximately proportional to $Z_h/D_0$. 
    Hence, a larger $Z_h$ or smaller $D_0$ requires a small value of $b_\chi$ to satisfy the AMS-02 and DAMPE B/C constraints. 

    \item  in the second region ($2\mev<m_\chi<0.1\gev$), we observe two distinct peaks in the B/C ratio spectrum due to DM-CRs collisions, as shown in Fig.~\ref{fig:realbtc}.  These peaks of $2\mev<m_\chi<100\mev$ shift to $E_k/n$ around $ \mathcal{O}(10^{2})\gev$, 
    but the AMS-02 errors at the $10\gev<E_k/n<100\gev$ region are relatively small. Furthermore, the systematic uncertainties induced by $D_0$ and $Z_h$ are significantly reduced 
    at $E_k/n > 10\gev$.
    Consequently, as we decrease the value of $m_\chi$ within this mass range, the upper limits on $b_\chi$ become more stringent, while the systematic uncertainties gradually shrink.

    \item in the third region ($m_\chi<2\mev$),   
    when the values of $m_\chi$ are less than $2\mev$, the DM-induced peaks of the CR spectra enter the region with $E_k/n>100\gev$, 
    where the error bars of AMS-02 and DAMPE measurements are the largest. On the other hand, the propagation uncertainties in this $E_k$ region are significantly reduced.  
    Therefore, unlike the other two mass regions, increasing the value of $m_\chi$ leads to more stringent limits on $b_\chi$, with the impact of propagation uncertainties being relatively small. For $m_\chi<100\kev$, the current B/C spectra measurements face challenges in probing DM-CR collisions, unless additional data points with $E_k/n>3\tev$ become available in the future. 
\end{enumerate}

Finally, we would like to examine the impact of $Z_h/D_0$ on the constraints imposed on $b_\chi$.  
When we change either $D_0$ or $Z_h$ within $1\sigma$ region, except for the $m_\chi<2\mev$ region, we observe that the 95\% upper limits of $b_\chi$ are significantly altered, typically by around one order of magnitude. Furthermore, when combining the AMS-02 and DAMPE B/C ratio data in the region of small $Z_h/D_0$ uncertainties, 
we obtain $b_\chi<\mathcal{O}(10^{-7})$ for $m_\chi\simeq 2\mev$.

\section{Summary and conclusions}

Considering DM-CR inelastic scattering, the high-energy CRs can be smashed by plentiful non-relativistic DM particles. 
Such collisions can significantly alter the CR energy spectra whose shapes differ from the standard DM-free propagation. 
Thus, based on the successful CR propagation model and precise CR measurements, 
the interactions between DM-CRs can be constrained. 
In this work, we propose a model that DM-CR interaction mimics proton-CR interaction and 
a constant $b_\chi$ is used to scale all proton-CR inelastic collision cross-sections to the corresponding DM-CR inelastic collision cross-sections.
By assuming that the final particle kinetic energy distributions in $\chi-$CRs and $p-$CRs are identical when 
the incoming kinetic energy of $\chi$ and $p$ are the same in the CR rest frame, we can study 
the CR spectra after DM-CR inelastic collisions.

We begin by exploring a simplified scenario using a toy model, where DM exclusively interacts with oxygen. The aim is to examine cascade productions resulting from DM-oxygen interactions. Our findings reveal that the B/C ratio spectrum remains unchanged for DM masses exceeding the proton mass ($m_\chi>m_p$). However, it might exhibit dual peaks for $m_\chi<m_p$. These patterns persist even when considering the full spectrum of DM-CR collisions, although tracing the cascade production in this more comprehensive scenario is more challenging.
Focusing on a specific example with a coupling strength of $b_{\chi}=0.1$ and a DM mass of $m_{\chi}=0.1~\gev$, where DM interacts solely with Oxygen, we observe a notable deviation in the B/C spectrum due to the DM-CR interaction, a peak observed between $0.1~\gev$ and $10~\gev$ experiences an enhancement of about 1.5 times. However, in a realistic model where DM collides with all CRs, this peak can be enhanced by up to twice its original value.

Once including the full DM-CRs collisions, we demonstrate the impact of propagation parameters $D_0$ and $Z_h$. 
By evaluating the Chi-squared from our predictions and measured B/C ratio spectra from AMS-02 and DAMPE, 
we find propagation uncertainties can degenerate with $b_\chi$ if $m_\chi>m_p$. 
However, it can be more tricky for $m_\chi<m_p$. 
Because the position of the peak in the B/C ratio varies from different DM particle masses, 
the peak shifts to higher energy regions if the DM mass gets lighter.
In the DM mass region $0.1\mev<m_\chi<1\mev$,
the $95\%$ upper limits of $b_\chi$ are less affected by propagation parameter $Z_h/D_0$.  
This is due to the fact that the DM-induced spectrum peaks are located at $E_k/n>100\gev$ region where 
the propagation uncertainties (see the blue shaded regions in Fig.~\ref{fig:best_ben_heavy}) are shrinking.

In summary, by combining the AMS-02 and DAMPE B/C ratio data, our result shows that 
$b_\chi$ shall be less than $\mathcal{O}(10^{-7})$ if $m_\chi\simeq 2\mev$, 
despite of a large propagation uncertainty $\mathcal{O}(10)$. 
For $m_\chi=100\kev$, the propagation uncertainties are smaller, 
but the upper limit becomes $b_\chi<\mathcal{O}(10^{-5})$.

\section*{Acknowledgements}
We thank Sujie Lin for discussing the \texttt{GALPROP} code. This work is supported by the National Key Research and Development Program of China (2022YFF0503304, 2020YFC2201600, 2018YFA0404504, 2018YFA0404601), the Ministry of Science and Technology of China (2020SKA0110402, 2020SKA0110401, 2020SKA0110100), the National Natural Science Foundation of China (11890691, 12205388, 12220101003), the CAS Project for Young Scientists in Basic Research (YSBR-061, YSBR-092), the China Manned Space Project with No. CMS-CSST-2021 (A02, A03, B01), the Major Key Project of PCL, and the 111 project (B20019).

\bibliographystyle{JHEP}
\bibliography{main}
\end{document}